\documentclass[a4paper,12pt]{JHEP3}

\usepackage{latexsym}
\usepackage{amsmath}
\usepackage{amsfonts}
\usepackage{amssymb}
\usepackage{dsfont}
\usepackage[dvips]{graphicx}

\newcommand{\hal}{{\textstyle\frac{1}{2}}}

\newcommand{\ca}[1]{{\cal{#1}}}

\newcommand{\unit}{\mathds{1}}  
\newcommand{\fr}[2]{{\textstyle{\frac{#1}{#2}}}}
\newcommand{\D}[1]{\dot{#1}}
\newcommand{\del}{\partial}
\newcommand{\vep}{\varepsilon}
\newcommand{\csep}{\setlength\arraycolsep{2pt}}
\newcommand{\ov}[1]{\overset{\mathrm{o}}{#1}}

\newcommand{\gym}{g_{_\textrm{YM}}}

\newcommand{\be}{\begin{equation}}
\newcommand{\ee}{\end{equation}}
\newcommand{\ben}{\begin{displaymath}}
\newcommand{\een}{\end{displaymath}}

\newcommand{\bea}{\begin{eqnarray}}
\newcommand{\eea}{\end{eqnarray}}

\newcommand{\ft}[2]{{\textstyle {\frac{#1}{#2}} }}

\makeatletter \@addtoreset{equation}{section} \makeatother

\author{Henning Samtleben and Robert Wimmer\\
Universit\'e de Lyon, Laboratoire de Physique,\\
Ecole Normale Sup\'erieure de Lyon,\\
46, all\'ee d'Italie, F-69364 Lyon Cedex 07, France\\
{\tt henning.samtleben, robert.wimmer\,\, @ens-lyon.fr}
}

\title{${\cal{N}}=8$ Superspace Constraints for Three-dimensional Gauge Theories}

\abstract{
We present a systematic analysis of the ${\cal{N}}=8$ superspace constraints 
in three space-time dimensions.
The general coupling between vector and scalar supermultiplets is encoded
in an $SO(8)$ tensor $W_{\!AB}$  which is a function of the matter fields and subject to a set of algebraic
and super-differential relations.
We show how the conformal BLG model as well as three-dimensional 
super Yang-Mills theory provide solutions to these constraints and
can both be formulated in this universal framework.
}


\preprint{ENSL-00439594}

\begin{document}   

\section{Introduction}

Highly supersymmetric three-dimensional gauge theories have
received tremendous attention over the last two years, 
in particular conformally symmetric matter Chern-Simons gauge theories. The origin of this  
interest was triggered by the formulation of the BLG-model 
\cite{Bagger:2007jr,Gustavsson:2007vu},  a non-trivially interacting $\ca{N}=8$ 
supersymmetric matter Chern-Simons gauge theory. 
It is an example of the 
sought-after theories describing the low energy dynamics of $M2$-branes and 
the conformally invariant fixed point of $\ca{N}=8$ SYM theory  \cite{Schwarz:2004yj}. Since then highly 
supersymmetric Chern-Simons gauge theories  
have been studied as examples of the  $AdS_4/CFT_3$-correspondence and as solvable idealizations of condensed matter 
systems at the conformal fixed point \cite{strings2009}. Progress has been made especially for $\ca{N}\leq 6$ 
supersymmetric models. However,  the $\ca{N}=8$ case, corresponding to $M2$-branes in maximally symmetric compactified 
M-theory,  remains notoriously intractable. The unitary BLG model is essentially 
unique with gauge group $SO(4)$ and arbitrary Chern-Simons level, whereas the $\ca{N}=6$ supersymmetric
$U(N)\times U(N)$ ABJM model \cite{Aharony:2008ug} has a proposed enhanced $\ca{N}=8$ supersymmetry for 
Chern-Simons levels $k=1,2$, but a manifest $\ca{N}=8$ supersymmetric formulation seems to be out of reach. 
It is generally accepted that these models are $CFT$'s due to the quantized nature of the CS-coupling,  
for an explicit two-loop confirmation see \cite{Akerblom:2009gx}. 
For both kind of models Higgs mechanisms have been introduced to study the flow to non-conformal 
SYM theories\cite{Mukhi:2008ux,Aharony:2008ug}.

Existing $\ca{N}=8$ superfield approaches
\cite{Bandos:2008df, Cederwall:2008vd} using Nambu-brackets and pure spinors
specifically describe the BLG model.
In the  work presented  here, 
we formulate and analyze the $\ca{N}=8$ superspace constraints for 
general three-dimensional gauge theories
which enables us to describe conformal Chern-Simons models and SYM theories on the same footing 
within a universal formalism. 
The matter sector
is  described by a real scalar superfield $\Phi^I$ transforming in the vector representation
of the $SO(8)$  $R$-symmetry group. The gauge sector is described by a vector superfield which is 
an $SO(8)$ singlet. 
These superfields are subject to appropriate constraints to restrict the field content and we study the 
possible couplings of the 
gauge and matter superfields. The set of theories which are allowed by the 
consistency conditions of the constraints 
can be parametrized by an antisymmetric $SO(8)$ tensor $W_{AB}$, which is a function of the matter 
superfields subject to 
the following concise $SO(8)$-projection conditions:
\begin{equation}
\label{eq:intro}
\nabla_{\alpha A} W_{BC}\;\Big|_{\bf 160_s} = 0\ , \qquad 
W_{IJ} \cdot \Phi_K\;\Big|_{\bf 160_v} = 0 \  ,\nonumber
\end{equation}
which will be explained in detail in the main text.
The  ${\cal{N}}=8$ superspace formulation implemented here is necessarily on-shell, so that pure 
superspace geometrical considerations of the multiplet 
structure determine the dynamics of the system in terms of superfield equations 
of motions. This is in analogy with the approach of 
\cite{Bagger:2007jr,Gustavsson:2007vu}, where the closure of the susy algebra led to the component field
e.o.m. However,
given a manifest super-covariant formulation 
the consistency checks of \cite{Bagger:2007jr,Gustavsson:2007vu} are automatically incorporated
in this framework and allow for a broader discussion of generalizations of the BLG-model. 

We give two classes of solutions to the above conditions which describe BLG-type 
conformal Chern-Simons gauge theories and maximal SYM theories, respectively. Lagrangian formulations are 
possible in terms of component fields here, and for the unitary BLG-model they require $SO(4)$ gauge group.
The existence of a Lagrangian description at the conformal fixed point is not guaranteed, though 
favorable conditions of $\mathbb{Z}_k$ orbifold $M$-theory compactifications make the existence of the
Lagrangian description by the 
ABJM models plausible \cite{jafferis}, but there is a hitch, in the case of the proposed $\ca{N}=8$ supersymmetry
with $k=1,2$ the theory is strongly coupled. Contrary to the four-dimensional $\ca{N}=4$ SYM theory 
there is no adjustable free parameter. In either case, existence of a strongly coupled Lagrangian or the lack of a Lagrangian description,  quantum theoretical considerations have to be done by other
means than perturbation theory within the models.

The superspace formulation that we present here provides a setting which allows the study of possible 
generalizations of BLG models and  the 
determination of quantum corrections  (to the e.o.m.) 
through symmetry 
considerations and by the rigidness of the $\ca{N}=8$ superspace, circumventing perturbation theory. We give an 
outline of possible strategies in the end of this paper.
The formulation of the dynamics in terms of superfield equations of motions carries enough information  
to investigate the 
moduli space of the theories as well as the possible chiral primary operators. Also the restrictions due the 
$\ca{N}=8$ superconformal symmetry as discussed in superspace in \cite{Park:1999cw} might be helpful for
further investigations. A big challenge in the $AdS_4/CFT_3$ correspondence remains  the understanding of 
the scaling of degrees of freedom with $N^{3/2}$  for the strongly coupled theory describing $N$ $M2$-branes
\cite{Berman:2007bv,Klebanov:2009sg}.

Finally we want to mention recent developments in $\ca{N}=8$ light-cone
superspace \cite{Ramond:2009hb,Belyaev:2009rj}. 

The paper is organized as follows. In section \ref{sec1} we review 
$\ca{N}=8$ superfields and the associated constraints for the free matter 
multiplet, the free Chern-Simons multiplet and discuss 
the minimal coupling of the matter multiplet to a free CS background.
In section~\ref{sec2} we introduce deformations of the free Chern-Simons constraint 
which are parametrized by an antisymmetric $SO(8)$ tensor $W_{AB}$.
We couple the matter sector to the gauge sector to obtain non-trivially interacting theories and 
show that consistency is equivalent to the above mentioned 
$SO(8)$-projection conditions for the tensor $W_{AB}$. 
We derive the explicit component equations of motion for general $W_{AB}$ 
and prove their equivalence to the superfield constraints.
In section~\ref{sec3} we give particular solutions to the above conditions, leading to BLG models 
and to a dual formulation of $\ca{N}=8$ SYM theories, respectively,
both embedded in the same superspace framework. 
We show how to explicitly re-dualize the SYM equations in $\ca{N}=8$ superspace.
In section \ref{sec4} we summarize 
our results and give an outlook on a number of future research directions. 
In appendix \ref{ap:WC} we analyze a weaker version of the superfield constraints
and derive the resulting multiplet structure and dynamics.

\section{Free CS multiplet and minimally coupled matter}
\label{sec1}

In this section we study the superspace description of the $\ca{N}=8$ super-multiplet
for free matter fields and for matter fields minimally coupled to a free Chern-Simons multiplet, thereby 
introducing the basic conventions and
methods used in this 
paper. The ${\cal{N}}=8$ superspace $\mathbb{R}^{2,1|16}$  is parametrized by 
coordinates $(x^{\alpha\beta},\theta^{\alpha A})$, $A=1,\ldots,8$, where the 
eight $\theta^{\alpha A}$ are real (Majorana) spinors in the 
$\bf{8_s}$ of the $SO(8)$ $R$-symmetry group and $x^{\alpha\beta}$ is a real symmetric 
matrix.\footnote{For more details regarding 
the notation see the appendix.} 
The susy covariant derivatives and the susy generators are 
given by the hermitian operators 
\begin{equation}
\label{eq:1.1}
D_{\alpha A}=\del_{\alpha A} + i\theta^\beta_A \del_{\alpha\beta}\ ,\qquad 
Q_{\alpha A}=\del_{\alpha A} - i\theta^\beta_A \del_{\alpha\beta}\ ,
\end{equation}
such that $\{ D_{\alpha A}, Q_{\beta B}\}=0$ and 
\begin{equation}
\label{eq:1.2}
\{ Q_{\alpha A}, Q_{\beta B}\} = - \{ D_{\alpha A}, D_{\beta B}\}=-2i\delta_{AB} \del_{\alpha\beta}\  .
\end{equation}
$SO(8)$ indices are raised/lowered with a Kronecker-delta and thus one does not have to pay special 
attention to their position. 
We will also use gauge covariant derivatives in superspace, which we introduce as follows:
\begin{equation}
\label{eq:1.3}
\nabla_{\alpha\beta}=\del_{\alpha\beta} + \mathcal{A}_{\alpha\beta} \ \ \textrm{and}\ \ \nabla_{\alpha A}=
D_{\alpha A} + \mathcal{A}_{\alpha A}\  . 
\end{equation}
When acting in complex bundles the physicality condition would be that the bosonic superspace connection 
$\mathcal{A}_{\alpha\beta}$ is anti-hermitian, while the 
fermionic one, $\mathcal{A}_{\alpha A}$, is hermitian, but we consider here real bundles and therefore 
the property under 
complex conjugation is the primary issue. To have the same conjugation property as for the  
differential operators 
we require that the 
bosonic superspace connection  $\mathcal{A}_{\alpha\beta}$ is real, while the 
fermionic one, $\mathcal{A}_{\alpha A}$, is imaginary.
Both connections carry a representation of the gauge symmetry structure group and complex conjugation has
to be defined accordingly. This and the action of the covariant derivatives on different fields will be 
discussed in detail when considering specific models.  

\subsection{The free matter multiplet}\label{sec:1.1}

{\subsubsection*{Superfield constraints}}
The $\ca{N}=8$ scalar multiplet consists of eight real scalars and eight 
Majorana-fermions $(\phi^I, \psi_{\alpha \dot{A}})$
in the $\bf{8_v}$ and  $\bf{8_c}$, respectively, of $SO(8)$. The free field equations are given by
\begin{equation}
\label{eq:1.3-2}
\Box \phi^I=0\ , \qquad \vep^{\beta\gamma}\del_{\alpha\beta}\psi_{\gamma \D{A}}=0\ ,
\end{equation}
where $\Box:=\del^{\alpha\beta}\del_{\alpha\beta}$.
The fields $\phi^I$ and consequently $\psi_{\alpha \D{A}}$ may carry an additional 
representation 
of some internal (global) symmetry group, which we do not indicate here but will be discussed in detail 
when we consider the interacting theories and systematically gauge these symmetries.

For finding superfields encoding this on-shell component multiplet
it is therefore natural to start with a real scalar superfield 
$\Phi^I$ in the $\bf{8_v}$ of $SO(8)$ 
(and in the same representation of a possible internal symmetry as $\phi^I$),
and impose necessary constraints to appropriately restrict  the component field content. 
At first order in $\theta^{\alpha A}$, this field contains components which transform as
$\bf{8_v}\otimes \bf{8_s}=\bf{8_c} \oplus \bf{56_c}$ under $SO(8)$.\footnote{
For details of $SO(8)$ representations and 
various $\Gamma$-matrix relations see the appendix \ref{ap:gam}. Decompositions of tensor products of 
$SO(8)$ representations can be computed with the program LiE \cite{LIE} or found in \cite{Slansky:1981yr}.} 
Comparing to the field content of
the component multiplet, it follows that one has to eliminate the unwanted component  
field in the $\bf{56_c}$.
In a susy covariant way this is achieved by imposing
\begin{equation}
\label{eq:1.4}
   D_{\alpha A}\Phi^I\big{|}_{{\bf{56_c}}}\overset{!}{=}0\ \quad \Longleftrightarrow\ \quad
    D_{\alpha A}\Phi^I =  \ft{1}{8} (\Gamma^I\bar{\Gamma}^J)_{AB}D_{\alpha B}\Phi^J\ .
\end{equation}
In  \cite{Cederwall:2008vd} a pure spinor superfield formulation of the BLG model was given and the equivalent to 
(\ref{eq:1.4}) was found as an invariance condition for the pure spinor wave-function. 

The constraint (\ref{eq:1.4}) implies the existence of a fermionic superfield $\Psi_{\alpha \D{A}}$
such that $D_{\alpha A}\Phi^I$  is  explicitly restricted to the $\bf{8_c}$:
\begin{equation}
\label{cons}
D_{\alpha A}\Phi^I =i \Gamma^I_{A\D{A}}\Psi_{\alpha\D{A}} \  \ ,
\end{equation}
and  for our purposes and in particular for applying the methods developed in 
\cite{Harnad:1984vk,Harnad:1985bc} it will be more convenient to work with this 
form of the constraint.
Equation (\ref{cons}) can be solved explicitly for $\Psi_{\alpha\D{A}}$ which by inserting gives 
back (\ref{eq:1.4}). This form of the constraint resembles the form of the ``super-embedding'' equation 
of \cite{Bandos:2008df}, where 
the BLG model was realized in terms of Nambu-brackets. The similarity will become more evident in the 
interacting case. 

The fermionic superfield $\Psi_{\alpha\D{A}} $ is not completely free, but is itself restricted
due to the  integrability condition of the constraint (\ref{cons}). With (\ref{eq:1.2}) this gives:
\begin{equation}
\label{ic}
2 \delta_{AB} \del_{\alpha\beta} \Phi^I=\Gamma^I_{A\D{A}}D_{\beta B}\Psi_{\alpha\D{A}}+
         \Gamma^I_{B\D{A}}D_{\alpha A}\Psi_{\beta\D{A}}\ ,
\end{equation}
which allows only the $(\bf{3},\bf{8_v})$ part of 
$D_{\alpha A}\Psi_{\beta \D{A}}$ to be nonzero, where the first entry 
refers to the  $SO(2,1)$ representation. We demonstrate here for once the procedure how we resolve such equations 
systematically. 
Decomposing $D_{\alpha A}\Psi_{\beta \D{A}}$ according to its irreducible representations
\begin{equation}
\label{eq:1.4-2}
\underbrace{D_{\alpha A}\Psi_{\beta \D{A}}}_{({\bf{2}}\otimes{\bf{2}},{\bf{8_s}}\otimes{\bf{8_c}})}                 
    =\Gamma^I_{A\D{A}}\underbrace{(\vep_{\alpha\beta}a^I
+a^I_{\alpha\beta})}_{({\bf{1}}\oplus{\bf{3}},{\bf{8_v}})}
    +\Gamma^{IJK}_{A\D{A}}\underbrace{(\vep_{\alpha\beta}b_{IJK}
          +b_{\alpha\beta IJK})}_{({\bf{1}}\oplus{\bf{3}},{\bf{56_v}})} \ ,         
\end{equation}
where the 3-form\footnote{The explicit form of a tensor in a representation of given dimension and 
the symmetries of these tensors are conveniently obtained via Young diagrams, see for example \cite{PaSGT}, though
their applicability is restricted for (special) orthogonal groups.} 
$b_{IJK}=b_{[IJK]}$ is the ${\bf{56_v}}$ and so is the $SO(2,1)$ vector $b_{\alpha\beta IJK}$. Inserting 
this decomposition into (\ref{ic}) shows that only the $(\bf{3},\bf{8_v})$  part $a^I_{\alpha\beta}$ can be non-zero 
and is given by the l.h.s.
The integrability condition (\ref{ic}) then  implies
\begin{equation}
\label{ic2}
D_{\alpha A}\Psi_{\beta \D{A}}= \Gamma^I_{A\D{A}}\del_{\alpha\beta} \Phi^I\ .
\end{equation}

The  constraint (\ref{cons}) and its integrability condition (\ref{ic2}) are 
the primary relations/conditions from which we derive all 
further consequences. 
From now on we will often refer to the constraint and its integrability condition as just the ``(superfield) constraints''.
Using (\ref{eq:1.2}) to express $x$-space 
derivatives in terms of superderivatives one  obtains that the superfields $\Phi^I$,  
$\Psi_{\alpha \D{A}}$ subject to the constraints (\ref{cons}), (\ref{ic2}) 
satisfy the free superfield e.o.m.
\csep
\begin{eqnarray}
\label{eq:sfe}
\vep^{\beta\gamma}\del_{\alpha\beta}\Psi_{\gamma \D{A}}&=&0 \;,\qquad
\Box\Phi^I= 0\ ,
\end{eqnarray}
where $\Box:=\del^{\alpha\beta}\del_{\alpha\beta}$. Thus the 
full superfields and therefore their lowest components 
$\phi^I:=\Phi^I|_{\theta=0}$, $\psi_{\alpha\D{A}}:=\Psi_{\alpha\D{A}}{|}_{\theta=0}$ 
(which are nonzero as we will see), 
satisfy the free e.o.m.~(\ref{eq:1.3-2}), as desired.  One could expect to get an additional condition from the integrability condition of (\ref{ic2}) but it is easy to see that it reduces to the superfield equations of motion (\ref{eq:sfe}). 

\subsubsection*{Superfield expansion}

Following \cite{Harnad:1984vk,Harnad:1985bc} we now derive recursion relations
which determine the $\theta$-expansion of the superfields. Defining the homogeneity operator
\begin{equation}
\label{eq:1.5}
\ca{R}:=\theta^{\alpha A}D_{\alpha A}=\theta^{\alpha A}\del_{\alpha A}\  ,
\end{equation}
which satisfies 
$\ca{R}(\theta^{\alpha_1 A_1}\ldots \theta^{\alpha_n A_n})=n\ \theta^{\alpha_1 A_1}\ldots \theta^{\alpha_n A_n}$, 
one obtains
by contracting the constraints  (\ref{cons}), (\ref{ic2}) 
with $\theta^{\alpha A}$ the \emph{recursion relations}
\csep
\begin{eqnarray}
\label{eq:rec}
\ca{R}\Phi^I&=&i\theta^{\alpha A} \Gamma^I_{A \D{A}}\Psi_{\alpha \D{A}}\;, \nonumber\\
\ca{R}\Psi_{\beta \D{A}}&=&\theta^{\alpha A} \Gamma^I_{A \D{A}}\del_{\alpha\beta}\Phi^I\ ,
\end{eqnarray}
which due to the property of $\ca{R}$  give the $(n+1)$'th order in $\theta$ of the superfields on the l.h.s 
in terms of the  $n$'th order of the superfields on the r.h.s.

The recursions (\ref{eq:rec}) determine the complete superfield expansion in terms of the lowest 
components $\phi^I$ and $\psi_{\alpha\D{A}}$, 
but without any further conditions on them and thus represent the non-dynamical part of the constraint equations. 
The resulting superfield expansion is:
\csep
\begin{eqnarray}
\label{eq:1.6}
\Phi^I&=&\phi^I+i \theta^{\alpha A}\Gamma^I_{A\D{A}}\psi_{\alpha \D{A}}
    +\fr{i}{2} \theta^{\alpha A} \theta^{\beta B}\Gamma^{IJ}_{AB}\del_{\alpha\beta}\phi^J+\ldots \;,\nonumber\\
\Psi_{\beta\D{A}}&=& \psi_{\beta\D{A}}+\theta^{\alpha A}\Gamma^I_{A\D{A}}\del_{\alpha\beta}\phi^I
  +\fr{i}{2} \theta^{\alpha A} \theta^{\gamma B}\Gamma^I_{A\D{A}}\Gamma^I_{B\D{B}}\del_{\alpha\beta}
     \psi_{\gamma\D{B}}+\ldots\;.
\end{eqnarray}
Given that the supersymmetry variation of a superfield $F$ is $\delta F=\epsilon^{\alpha A}Q_{\alpha A} F$ 
one obtains from (\ref{eq:1.1}) the following transformations for the component fields:
\begin{equation}
\label{eq:1.7}
\delta\phi^I=i\epsilon^{\alpha A}\Gamma^I_{A\D{A}}\psi_{\alpha\D{A}}\ ,\qquad 
\delta\psi_{\beta \D{A}}=\epsilon^{\alpha A}\Gamma^I_{A\D{A}}\del_{\alpha\beta}\phi^I\ ,
\end{equation}
which by construction are symmetries of the 
e.o.m.~(\ref{eq:1.3-2}). 

Concluding, we have shown that the superfield constraints (\ref{cons}), (\ref{ic2}) imply a superfield expansion exclusively
in terms of the component multiplet ($\phi^I,\psi_{\alpha \D{A}}$)  with the supersymmetry transformations (\ref{eq:1.7}).
Moreover, these superfields satisfy the free superfield e.o.m.~(\ref{eq:sfe}) and 
thus the component fields satisfy the free e.o.m.~(\ref{eq:1.3-2}). 
In the rest of this section, we will show that vice versa
the on-shell component fields define superfields which satisfy the constraints (\ref{cons}), (\ref{ic2}) so  
that these two descriptions are completely equivalent.
In particular, the constraints (\ref{cons}), (\ref{ic2}) do not imply any further restrictions 
on the components.

\subsubsection*{Equivalence to component e.o.m.} 
We now start from 
the on-shell component multiplet ($\phi^I,\psi_{\alpha \D{A}}$) ,  which
is assumed to satisfy the free e.o.m (\ref{eq:1.3-2}), 
which are supersymmetric under the transformations (\ref{eq:1.7}),
and show that this defines superfields satisfying the constraints (\ref{cons}), (\ref{ic2}). 

{\bf{Susy covariance.}} 
We use the recursion relations (\ref{eq:rec}) to define 
superfields out of the component multiplet ($\phi^I,\psi_{\alpha \D{A}}$).
For the 
first few terms in the $\theta$-expansion (\ref{eq:1.6}) 
we have already shown that the component supersymmetry transformations  (\ref{eq:1.7})
can be written as $\delta\Phi^I=\epsilon Q\Phi^I$, $\delta\Psi_{\alpha\D{A}}=\epsilon Q\Psi_{\alpha\D{A}}$, with 
$Q_{\alpha A}$ given in (\ref{eq:1.1}).
The recursion relations  (\ref{eq:rec}) are not susy covariant and one has to check explicitly if they define a 
consistent superfield\footnote{ 
This  is a complementary approach for finding the correct superfield 
constraints for a given multiplet with susy transformations (\ref{eq:1.7}) which upon comparing with 
(\ref{eq:rec}) define the
recursion relations so that the superfield expansion is generated by consecutive susy transformations.}, 
i.e.\ that susy transformed superfields satisfy the same 
recursion relations. 

Acting with $\epsilon^{\alpha A}Q_{\alpha A}$ on the recursion relations (\ref{eq:rec}) one obtains
\csep
\begin{eqnarray}
   \label{eq:1.9}
   \ca{R}\delta\Phi^I&=&i\theta^{\alpha A} \Gamma^I_{A \D{A}}\delta\Psi_{\alpha \D{A}} 
           -\epsilon^{\alpha A}[D_{\alpha A}\Phi^I  - i \Gamma^I_{A\D{A}}\Psi_{\alpha\D{A}} ]\ ,\nonumber\\
\ca{R}\delta\Psi_{\beta \D{A}}&=&\theta^{\alpha A} \Gamma^I_{A \D{A}}\del_{\alpha\beta}\delta\Phi^I 
         -\epsilon^{\alpha A}[D_{\alpha A}\Psi_{\beta \D{A}} - \Gamma^I_{A\D{A}}\del_{\alpha\beta} \Phi^I]\  .
 \end{eqnarray}
Thus the susy variations satisfy the same recursions as the original fields \emph{iff} 
the superfield constraints (\ref{cons}), (\ref{ic2}) are satisfied. To show that the 
component e.o.m.\ imply these constraints we first prove that they imply the full superfield e.o.m. 

{\bf{Superfield e.o.m.}} To zeroth order in $\theta$, the superfields 
equal the components  ($\phi^I,\psi_{\alpha \D{A}}$) and 
thus per construction satisfy the e.o.m. To show that this implies that they are satisfied in all orders 
in $\theta$ we derive a 
recursive system for the superfield e.o.m.,
\csep
\begin{eqnarray}
\label{eq:1.10}
\ca{E}_{\alpha\D{A}}&:=&\vep^{\beta\gamma}\del_{\alpha\beta}\Psi_{\gamma\D{A}}\ \ ,\qquad
\ca{E}^I:=\Box\Phi^I\ \ .
\end{eqnarray}
Using exclusively the recursion relations (\ref{eq:rec}) one obtains\footnote{Here and on many 
other occasions we use the fact that the total antisymmetrization of three spinor indices, 
which take two values, gives zero.}
\csep 
\begin{eqnarray}
\label{eq:1.11}
\ca{R}\ca{E}_{\alpha\D{A}}&=&-\theta^{\beta A}\Gamma^I_{A\D{A}}\vep_{\beta\alpha}\ca{E}^I\ \ ,\nonumber\\
\ca{R}\ca{E}^I&=&
     -i\theta^{\alpha A}\Gamma^I_{A\D{A}}\vep^{\beta\gamma}\nabla_{\alpha\beta}\ca{E}_{\gamma \D{A}}\ \ .
\end{eqnarray}
As to lowest order the e.o.m.\ are satisfied, i.e.\ 
$\ca{E}_{\alpha\D{A}}|_{\theta=0}=\ca{E}^I|_{\theta=0}=0$, these recursions imply 
that $\ca{E}_{\alpha\D{A}}$, $\ca{E}^I$ vanish to
all orders. Thus the component e.o.m.~(\ref{eq:1.3-2}) imply the superfield e.o.m.~(\ref{eq:sfe}) for the superfields defined by~(\ref{eq:rec}).

{\bf{Constraints.}} In the last step we show that the superfield e.o.m.~(\ref{eq:sfe}) imply the constraint equations 
(\ref{cons}), (\ref{ic2}). To this end we introduce the abbreviations
\csep 
\begin{eqnarray}
\label{eq:1.12}
\ca{C}^I_{\alpha A}&=& D_{\alpha A}\Phi^I  - i\ \Gamma^I_{A\D{A}}\Psi_{\alpha\D{A}}\ \ ,\nonumber\\
\ca{C}_{\alpha\beta A \D{A}}&=&D_{\alpha A}\Psi_{\beta \D{A}} 
- \Gamma^I_{A\D{A}}\del_{\alpha\beta} \Phi^I\ \ . 
\end{eqnarray}
Using the recursion relation (\ref{eq:rec}) one obtains the following recursions 
for the constraints $ \ca{C}^I_{\alpha A}$ and $ \ca{C}_{\alpha\beta A \D{A}}$:
\csep
\begin{eqnarray}
\label{eq:1.13}
(1+\ca{R}) \,\ca{C}^I_{\alpha A}&=&
    i\theta^{\beta B}\Gamma^I_{B\D{A}}\ca{C}_{\alpha\beta A \D{A}}\ \ ,\nonumber\\
(1+\ca{R}) \,\ca{C}_{\alpha\beta A \D{A}}&=&
   -\theta^{\gamma B}\Gamma^I_{B\D{A}}\del_{\beta\gamma}\ca{C}^I_{\alpha A}\  \ ,
\end{eqnarray}
where in the second relation we used the fermionic superfield e.o.m.~(\ref{eq:sfe}).
These recursions imply that the constraints $ \ca{C}^I_{\alpha A}$ and $ \ca{C}_{\alpha\beta A \D{A}}$ vanish in all
orders in $\theta$. We thus have proved that the on-shell multiplet  ($\phi^I,\psi_{\alpha \D{A}}$) 
with equations of motion (\ref{eq:1.3-2}) is completely equivalent to 
the superfields ($\Phi^I,\Psi_{\alpha \D{A}}$) satisfying the constraints  (\ref{cons}), (\ref{ic2}).

\subsection{Free Chern-Simons multiplet}
\label{sec:fcs}

In general, we will be interested in theories whose matter content is given by a number
of scalar super-multiplets. At the linearized level, such theories are described by $N$~superfields 
$\Phi^{Ia}, \Psi_{\alpha\D{A}}^a$, subject to the constraint (\ref{cons}), where the additional index
$a=1, \dots N$, labels the different super-multiplets.
The obvious global symmetry group (besides the $SO(8)$ $R$-symmetry, 
which we will not gauge)
of the  system is 
$GL(N,\mathbb{R}) \ltimes \mathbb{T}(8N)$ acting as 
\begin{equation}
\label{eq:1.14}
\delta\Phi^I=\Lambda\cdot\Phi^I+ C^I\ ,
\qquad \delta\Psi_{\alpha\D{A}}=\Lambda\cdot\Psi_{\alpha\D{A}}\ ,
\end{equation}
with a matrix $\Lambda\in\mathfrak{gl}(N,\mathbb{R})$ (where we have suppressed the explicit indices $a$),
which are obviously symmetries of (\ref{cons}).
The shifts ${\mathbb{T}}(8N)$ act exclusively on the scalars~$\Phi^I$.\footnote{The component 
field equations (\ref{eq:1.3-2}) would allow also for global shifts 
$\delta\psi_{\alpha\D{A}}=\zeta_{\alpha\D{A}}$ of the fermionic component field and thus of the superfield 
$\Psi_{\alpha\D{A}}$. 
In view of the superfield expansion (\ref{eq:1.6}) this would imply a corresponding $\theta$-dependent shifts 
$\delta\Phi^I=C^I+i\theta^{\alpha A}\Gamma^I_{A \D{A}}\zeta_{\alpha\D{A}}$ in the bosonic superfield
$\Phi^I$ and represent a more involved symmetry of the constraints (\ref{cons}), (\ref{ic2}). We do not 
consider this possibility here.}

In the interacting theories, a subset of these symmetries will be gauged by selecting a subalgebra
$\mathfrak{g}$
\setlength\arraycolsep{2pt}
\begin{eqnarray}
\label{eq:1.16}
\langle T_M\rangle&=&{\mathfrak{g}} \subset 
       {\mathfrak{gl}}(N,\mathbb{R})\oplus_s{\mathfrak{t}(8N)}\ ,\nonumber\\
\ [T_M,T_N]&=&f_{MN}^{\ \ \ K}\ T_{K} \;,
\end{eqnarray}
spanned by generators $T_M$\,.
Choosing $\mathfrak{g}$ to have non-trivial intersection with $\mathfrak{t}(8N)$  
a priori breaks the $SO(8)$ $R$-symmetry. 
The corresponding gauge superfields appearing in the covariant derivatives (\ref{eq:1.3}) 
are given by
\begin{equation}
\label{eq:1.15}
\ca{A}_{\alpha A}=\ca{A}^M_{\alpha A}iT_M\ ,\qquad \ca{A}_{\alpha \beta}=\ca{A}^M_{\alpha \beta}T_M\ .
\end{equation}
Assuming a real representations for the generators $T_M$, this gives the right conjugation property
for real $\ca{A}^M_{\alpha A}$ and $\ca{A}_{\alpha \beta}^M$, as 
defined below (\ref{eq:1.3}).  

Note that at this stage we do not encounter three algebras as introduced in 
\cite{Bagger:2006sk,Bagger:2007jr,Gustavsson:2007vu}.  
We will see in later sections how the
defining relation of these three algebras, the \emph{fundamental identity} for 
a rank four tensor, is a natural consequence 
for conformal models based on Lie algebras.  

Introducing the gauge parameter field $\Omega=\Omega^M T_M$, the local versions of (\ref{eq:1.14}) and the gauge 
transformations of the gauge fields can be compactly written as 
\begin{eqnarray}
\label{eq:1.21}
\delta\Phi^I=\Omega\cdot\Phi^I &,&\quad 
            \delta\Psi_{\alpha\D{A}} = \Omega\cdot\Psi_{\alpha\D{A}}\ ,\nonumber\\
\delta\ca{A}_{\alpha A} = -\nabla_{\alpha A}\Omega&,&\quad 
\delta \ca{A}_{\alpha \beta} = -\nabla_{\alpha\beta}\Omega \ , 
\end{eqnarray}
where the gauge fields transform in the adjoint of (\ref{eq:1.16}) and the matter superfields
now transform in some representation of the gauge algebra which is indicated by the dot.

The field strengths are given in the usual way through (anti)commutators of the connections minus 
torsion terms, i.e.\ 
\setlength\arraycolsep{2pt}
\begin{eqnarray}
\label{eq:1.24}
\ca{F}_{\alpha A, \beta B}&=&
     \{\nabla_{\alpha A},\nabla_{\beta B}\}-2i \delta_{AB}\nabla_{\alpha\beta}\ , \nonumber\\
\ca{F}_{\alpha \beta, \gamma\delta}&=&[\nabla_{\alpha\beta},\nabla_{\gamma\delta}]\ , \nonumber\\
\ca{F}_{\alpha \beta, \gamma C}&=&[\nabla_{\alpha\beta},\nabla_{\gamma C}] \ .
\end{eqnarray}

\subsubsection*{Free CS superfield constraints} 
The gauge superfields 
($\ca{A}_{\alpha A}$, $\ca{A}_{\alpha\beta}$) contain
way to many component fields and one has to impose constraints to obtain a physically meaningful multiplet. 
It has turned out to be promising to impose (partial) flatness conditions
on the bi-spinor field strength, here  $\ca{F}_{\alpha A, \beta B}$,  to eliminate unphysical degrees of freedom 
\cite{Grimm:1977xp, Sohnius:1978wk, Witten:1978xx, Witten:1985nt}. In many cases this 
corresponds to an underlying geometric structure 
of twistors and pure spinors \cite{ Witten:1978xx, Witten:1985nt,Harnad:1988rs,Howe:1991mf}.   

The bi-spinor field strength contains the representations
\begin{equation}
 \label{convent}
 \ca{F}_{\alpha A, \beta B}\sim (({\bf{2}},{\bf{8_s}})\otimes ({\bf{2}},{\bf{8_s}}))_{\mathrm{sym}} =
  ({\bf 3},{\bf 1})\oplus ({\bf 1},{\bf 28}) \oplus({\bf 3},{\bf 35}) .
\end{equation}
The $({\bf 3},{\bf 1})$ part corresponds to a second component vector field in the superfield expansion of
$\ca{A}_{\alpha A}$ with the same gauge-transformation as the lowest component of  $\ca{A}_{\alpha\beta}$. 
Following the standard approach, see for example \cite{Gates:1979wg,Gates:1983nr}, 
we will set this part to zero as the so-called ``conventional constraint'',  
which in particular eliminates the additional component vector field.
Putting further constraints on $\ca{F}_{\alpha A, \beta B}$, in contrast, will not only eliminate component fields,
but also induce (partial) equations of motion for the remaining fields. We shall analyze this in more detail in this paper.

Since we are interested here in the free multiplet we impose in this section
a constraint which is  rather strong in three dimensions and require the 
entire $\ca{F}_{\alpha A, \beta B}$ to vanishes. Relaxations of this constraint will be discussed when we 
consider non-minimally interacting 
theories. Thus, for this section we set
\begin{equation}
\label{eq:1.25}
\ca{F}_{\alpha A, \beta B}\overset{!}{=}0 \ \quad \Longleftrightarrow \ \quad 
         \{\nabla_{\alpha A},\nabla_{\beta B}\}=2 i \delta_{AB}\nabla_{\alpha\beta} 
         \;.
\end{equation}

As in the case of the matter superfield constraint (\ref{cons}) the right r.h.s.\! of (\ref{eq:1.25}) is not completely free but 
has to satisfy certain conditions so that it factorizes into an anti-commutator. The analogon to the integrability condition 
(\ref{ic2}) are the Bianchi identities, which are simply obtained from the super-Jacobi identities 
for the covariant derivatives:\footnote{The exponent $\pi$ in the second 
identity counts the cyclic permutations where (anti)commutators are distributed correspondingly to the occurrence of
bosonic/fermionic connections} 
\begin{eqnarray}
\label{eq:BI}
\sum_{\textrm{cyclic}}[\nabla_{\alpha A},\{\nabla_{\beta B},\nabla_{\gamma C}\}]\equiv 0 &,&\qquad
\sum_{\textrm{cyclic}}(-1)^{\pi}\{\nabla_{\alpha A},[\nabla_{\beta B},\nabla_{\gamma\delta}]\}\equiv 0\ ,\nonumber\\
\sum_{\textrm{cyclic}}[\nabla_{\rho A},[\nabla_{\alpha\beta},\nabla_{\gamma\delta}]]\equiv 0\quad   &,&\qquad 
\sum_{\textrm{cyclic}}[\nabla_{\alpha\beta},[\nabla_{\gamma\delta},\nabla_{\rho\sigma}]]\equiv 0\ .
\end{eqnarray}
First of all, these identities imply nontrivial conditions  
in case (\ref{eq:1.25}) appears, i.e.\ for the first and second identity with three fermionic and two fermionic 
covariant derivatives, respectively.
In these cases one obtains:
\setlength\arraycolsep{2pt}
\begin{eqnarray}
\label{eq:1.26}
\delta_{AB}\ca{F}_{\alpha\beta,\gamma C}+\delta_{AC}\ca{F}_{\alpha\gamma,\beta B}
             +\delta_{BC}\ca{F}_{\beta\gamma,\alpha A}&=&0\ , \nonumber\\
\nabla_{\alpha A}\ca{F}_{\gamma\delta,\beta B}+ \nabla_{\beta B}\ca{F}_{\gamma\delta,\alpha A}
 &=& 2 i \delta_{AB}\ca{F}_{\gamma\delta,\alpha\beta}\ .
\end{eqnarray}
Decomposing the two equations analogously to (\ref{eq:1.4-2}) into irreducible representations  of $SO(2,1)$ and $SO(8)$,   
one finds that the two Bianchi identities
imply that also the other two components of the super field strength vanish, i.e.\
\setlength\arraycolsep{2pt}
\begin{eqnarray}
\label{eq:1.27}
\ca{F}_{\alpha\beta,\gamma C}&=&0=
\ca{F}_{\alpha\beta,\gamma\delta}\ .
\end{eqnarray}
With these strong equations for the commutators/field strengths the other two
Bianchi identities in (\ref{eq:BI}) are identically fulfilled and do not impose further conditions. 
The second equation in (\ref{eq:1.27}), which follows with the help of the first one, 
is the free Chern-Simons superfield equation of motion. 
To see what this implies at the level of component fields we again 
follow the strategy of~\cite{Harnad:1984vk,Harnad:1985bc} to obtain the superfield expansion. 

{\subsubsection*{Superfield expansion}} 
To eliminate the gauge degrees of freedom in the gauge superfields and to
be able to apply the same recursive method as in (\ref{eq:rec}) one imposes the ``transverse'' gauge 
\cite{Harnad:1984vk} 
on the fermionic
gauge superfields,
\begin{equation}
\label{eq:1.28}
\theta^{\alpha A}\ca{A}_{\alpha A}=0\quad \Longrightarrow\quad 
\ca{R}= \theta^{\alpha A}\nabla_{\alpha A}\ .
\end{equation}
This fixes the gauge freedom (\ref{eq:1.21}) up to pure $x$-space dependent gauge transformations 
and is thus a kind of
WZ-gauge. Moreover, it allows to write the recursion operator $\ca{R}$ (\ref{eq:1.5}) in a covariant form. 
Therefore, contracting the constraint 
(\ref{eq:1.25}) and the first Bianchi identity (\ref{eq:1.27}) with $\theta^{\gamma C}$ one obtains the 
recursion relations
\csep
\begin{eqnarray}
\label{eq:1.29}
(1+\ca{R})\ \ca{A}_{\beta B}&=&2 i \theta^{\alpha}_{ B}\ca{A}_{\alpha\beta}\ ,\nonumber\\
  \ca{R}\ \ca{A}_{\alpha\beta}&=&0\ .
\end{eqnarray}
This gives the rather trivial superfield expansions,
\begin{equation}
\label{eq:1.30}
\ca{A}_{\alpha A}= i\theta^{\beta}_{A} A_{\alpha\beta}\ ,\qquad               
 \ca{A}_{\alpha\beta}=A_{\alpha\beta}\ ,
\end{equation}
where the lowest component $A_{\alpha\beta}:=\ca{A}_{\alpha\beta}|_{\theta=0}$ is the vector field in $x$-space. 
The condition due to the second Bianchi identity  in (\ref{eq:1.27}) thus implies the component field equations
\begin{equation}
\label{eq:1.31}
F_{\alpha\beta, \gamma\delta}=0\ ,
\end{equation}
which is the free Chern-Simons e.o.m. Consequently,  
the multiplet associated with the constraint (\ref{eq:1.25}) contains a single component field, 
the vector field $A_{\alpha\beta}$, which
describes a flat connection
and therefore has no
local degrees of freedom.

{\subsubsection*{Equivalence to component e.o.m.}} 
To prove that a 
component vector field $A_{\alpha\beta}$, satisfying (\ref{eq:1.31}) is equivalent to the full constraint 
(\ref{eq:1.25}) is trivial in this case. Adopting the superfield expansions (\ref{eq:1.30}) one  immediately sees that these
superfields satisfy the constraint (\ref{eq:1.25}) and the Bianchi identities (\ref{eq:1.27}) 
due to the component field e.o.m.~(\ref{eq:1.31}).
Nevertheless, we consider the susy-covariance of the recursion relations and the susy-transformations 
of the component field. Defining the superfield transformations as before, i.e.\ 
$\delta\ca{A}_{\alpha A}=\epsilon Q\ca{A}_{\alpha A}$ and 
$\delta\ca{A}_{\alpha\beta}=\epsilon Q\ca{A}_{\alpha\beta}$, and acting with $\epsilon Q$ on the recursions 
(\ref{eq:1.29}) one finds:
\csep
\begin{eqnarray}
\label{eq:1.32}
(1+\ca{R})\,\delta\ca{A}_{\alpha A}&=&2i\theta^{\beta}_B\delta\ca{A}_{\alpha\beta} 
                 -\epsilon^{\beta B}\ca{F}_{\alpha A,\beta B}
                   -\nabla_{\alpha A}\Lambda\;,\nonumber\\ 
\ca{R}\delta\ca{A}_{\alpha\beta}&=&0  
                     +\epsilon^{\gamma C}\ca{F}_{\alpha\beta,\gamma C} -\nabla_{\alpha\beta}\Lambda\ , 
\end{eqnarray}
where  in both cases the last term is a field dependent supergauge transformation with the gauge parameter field
\begin{equation}
\label{eq:1.33}
\Lambda=\epsilon^{\alpha A}\ca{A}_{\alpha A}=i\epsilon^{\alpha A}\theta^{\beta}_{A}A_{\alpha\beta}\ .
\end{equation}
Thus up to the constraint (\ref{eq:1.25}) and the first Bianchi identity in (\ref{eq:1.27}) the recursion relations are 
susy covariant modulo 
field dependent gauge transformations. The occurrence of the field dependent gauge transformation is not surprising since 
the ``transverse'' gauge (\ref{eq:1.28}) is not susy covariant.  In the same fashion, using
the (component) field e.o.m.~(\ref{eq:1.31}) and in view of the superfield expansion (\ref{eq:1.30})
one obtains for the supersymmetry transformations
\csep
\begin{eqnarray}
\label{eq:1.34}
\delta\ca{A}_{\alpha A}&=&\epsilon^{\gamma C}Q_{\gamma C} \ca{A}_{\alpha A}
       =:i\theta^{\beta}_{A}\ \delta A_{\alpha\beta}=-\nabla_{\alpha A}\Lambda\ ,\nonumber\\
 \delta\ca{A}_{\alpha \beta}&=&\epsilon^{\gamma C}Q_{\gamma C}\ca{A}_{\alpha \beta} 
       =:\ \delta A_{\alpha\beta}\ =-\nabla_{\alpha \beta}\Lambda\ .
\end{eqnarray}
Thus the susy transformations of the superfields are (on-shell) pure gauge transformations with the field dependent 
parameter $\Lambda$  (\ref{eq:1.33}). These gauge transformations do not have a component in the appropriate order 
of $\theta$ such that the supersymmetry transformation of the component field in (\ref{eq:1.34}) is just
\begin{equation}
\label{eq:1.35}
\delta A_{\alpha\beta}=0\ .
\end{equation}
The curious fact for the free Chern-Simons case, that a multiplet with a single component field, $A_{\alpha\beta}$, is 
nevertheless (on-shell) supersymmetric
was discussed in \cite{Schwarz:2004yj}. 
Here we obtain the same result in a super-covariant way.  

\subsection{Minimal Coupling of matter to free CS} 
\label{sec:mc}

We now covariantize the procedure of section \ref{sec:1.1} 
by minimally coupling  the matter superfields 
to gauge superfields subject to the constraint (\ref{eq:1.25}).
It therefore describes matter fields minimally coupled to a free CS background (without backreaction).
Given that the gauge field remains a flat connection 
this seems to be trivial, but it sets the formalism for  the next section,
where we consider non-linear deformations which lead to a non-trivially coupled system.

{\subsubsection*{Superfield Constraints}} The covariantized constraint (\ref{cons}) with minimal coupling is
\begin{equation}
\label{c2}
\nabla_{\alpha A}\Phi^I=i\Gamma^I_{A \D{A}}\Psi_{\alpha \D{A}} \ .
\end{equation}
Using the gauge field constraint (\ref{eq:1.25}) and the Bianchi identities (\ref{eq:1.27}), the integrability condition of
(\ref{c2}) reduces to
\begin{equation}
\label{IC}
\nabla_{\alpha A}\Psi_{\beta\D{A}}=\Gamma^I_{A\D{A}}\nabla_{\alpha\beta}\Phi^I\ .
\end{equation}
Further, using the gauge field constraint (\ref{eq:1.25}) to express $\nabla_{\alpha \beta}$ in terms
of superderivatives and 
the Bianchi identities (\ref{eq:1.27}) together with 
(\ref{c2}), (\ref{IC}), the superfield e.o.m.\ compute to
\csep
\begin{eqnarray}
\label{eq:sfeom}
\vep^{\beta\gamma}\nabla_{\alpha\beta}\Psi_{\gamma\D{A}}&=&0 \,, \qquad
\nabla^2\Phi^I=0\ ,
\end{eqnarray}
where  $\nabla^2=\nabla^{\alpha\beta}\nabla_{\alpha\beta}$.

{\subsubsection*{Superfield expansion}} To obtain the superfield expansion 
we again impose the ``transverse'' gauge (\ref{eq:1.28}). Contracting the 
constraints (\ref{c2}), (\ref{IC}) with $\theta^{\alpha A}$ one obtains the recursion 
relations
\csep
\begin{eqnarray}
\label{eq:1.36}
\ca{R}\Phi^I&=&i\theta^{\alpha A} \Gamma^I_{A\D{A}}\Psi_{\alpha\D{A}}\ ,\nonumber\\
\ca{R}\Psi_{\beta\D{A}}&=&\theta^{\alpha A}\Gamma^I_{A\D{A}}\nabla_{\alpha\beta}\Phi^I\ .
\end{eqnarray}
This again defines the superfield expansion in terms of the lowest components $\phi^I=\Phi^I|_{\theta=0}$ and 
$\psi_{\alpha\D{A}}=\Psi_{\alpha\D{A}}|_{\theta=0}$, where things considerable simplify due to the 
fact that for the free Chern-Simons
multiplet in the ``transverse'' gauge (\ref{eq:1.30}) one has 
\begin{equation}
\label{eq:1.36-2}
     \nabla_{\alpha\beta}=\ov{\nabla}_{\alpha\beta}:=\del_{\alpha\beta}+A_{\alpha\beta}\ , 
\end{equation}
i.e.\ only the lowest component of the vector superfield is present in the super-connection  $\nabla_{\alpha\beta}$.\footnote{
By `$\ov{\phantom{x}}$' we generically denote the lowest component of
a superfield: $\ov{\Phi}:=\Phi|_{\theta=0}=\phi$, etc.
}
Hence the superfield expansion is given by
\begin{eqnarray}
\label{eq:1.37}
\Phi^I&=&\phi^I+i \theta^{\alpha A}\Gamma^I_{A\D{A}}\psi_{\alpha \D{A}}
    +\fr{i}{2} \theta^{\alpha A} \theta^{\beta B}\Gamma^{IJ}_{AB}\ov{\nabla}_{\alpha\beta}\phi^J
           +\ldots\ , \nonumber\\
\Psi_{\beta\D{A}}&=& \psi_{\beta\D{A}}+\theta^{\alpha A}\Gamma^I_{A\D{A}}
           \ov{\nabla}_{\alpha\beta}\phi^I
  +\fr{i}{2} \theta^{\alpha A} \theta^{\gamma B}\Gamma^I_{A\D{A}}\Gamma^I_{B\D{B}}
          \ov{\nabla}_{\alpha\beta} \psi_{\gamma\D{B}}+\ldots ,
\end{eqnarray}
and therefore the lowest components of the superfield e.o.m.~(\ref{eq:sfeom}) imply the corresponding e.o.m for the
component fields $\phi^I$ and $\psi_{\alpha\D{A}}$. We have thus shown that the constraints and integrability 
conditions/Bianchi identities (\ref{eq:1.25}), (\ref{eq:1.27}), (\ref{c2}), (\ref{IC}) give a minimally coupled Chern-Simons multiplet 
$(A_{\alpha\beta},\phi^I,\psi_{\alpha\D{A}})$ with the e.o.m.
\begin{eqnarray}
\label{eq:1.38}
F_{\alpha\beta,\gamma\delta}=0\;,\qquad
\ov{\nabla}{}^2 \,\phi^I=0\;,\qquad
  \vep^{\beta\gamma}\ov{\nabla}_{\alpha\beta}\psi_{\gamma\D{A}}=0 \ .
\end{eqnarray}
The supersymmetry transformations of the matter multiplet are obtained from the superfield expansion (\ref{eq:1.37}) in the usual way,
\begin{equation}
\label{eq:1.39}
\delta\Phi^I=\epsilon^{\alpha A}Q_{\alpha A}\Phi^I
     =:(\delta\phi^I+i\theta^{\alpha A}\Gamma^I_{A\D{A}}\delta\psi_{\alpha\D{A}}\ldots) +\Lambda\cdot\Phi^I\ ,
\end{equation}
where as in the case of the gauge multiplet we obtain the component field transformations modulo a compensating gauge 
transformation with the same gauge parameter $\Lambda$ (\ref{eq:1.33}). The resulting supersymmetry transformations are then
\csep
\begin{eqnarray}
\label{eq:1.40}
&&\delta\phi^I=i\epsilon^{\alpha A}\Gamma^I_{A\D{A}}\psi_{\alpha\D{A}}\ ,\ 
  \delta\psi_{\alpha\D{A}}=\epsilon^{\beta A}\Gamma^I_{A\D{A}}\ov{\nabla}_{\alpha\beta}\phi^I\ ,\nonumber\\
&&\delta A_{\alpha\beta}=0\ ,
\end{eqnarray}
where for completeness we have rewritten the  transformation of the gauge field (\ref{eq:1.35}). These 
supersymmetry transformations again resemble the recursion relations (\ref{eq:1.36}), (\ref{eq:1.29}) of 
the associated superfields. 

{\subsubsection*{Equivalence to component e.o.m.}} 
We have already shown that the 
component vector field $A_{\alpha\beta}$ subject to the 
free Chern-Simons e.o.m.~(\ref{eq:1.31}) is equivalent to the gauge 
field constraint (\ref{eq:1.25}) and its Bianchi identities  (\ref{eq:1.27}). 
What remains to be shown is that the same is true for the matter multiplet. Again we start from the 
multiplet ($\phi^I$,$\psi_{\alpha\D{A}}$), satisfying the e.o.m.~(\ref{eq:1.38}) and construct superfields out of 
it according to the recursions (\ref{eq:1.36}). It is convenient to introduce again the constraint functions
\csep
\begin{eqnarray}
\label{eq:1.41}
\ca{C}^I_{\alpha A}&:=&\nabla_{\alpha A}\Phi^I-i\Gamma^I_{A\D{A}}\Psi_{\alpha \D{A}}\ , \nonumber\\
\ca{C}_{\alpha\beta A\D{A}}&:=&
      \nabla_{\alpha A}\Psi_{\beta\D{A}}-\Gamma^I_{A\D{A}}\nabla_{\alpha\beta}\Phi^I\ ,
\end{eqnarray}
where we used the same symbols as for the free matter multiplet, which now encode the minimally coupled constraints (\ref{c2}), (\ref{IC}) (but this should not lead to any confusion).   
Acting with $\epsilon^{\alpha A}Q_{\alpha A}$ on the recursions
(\ref{eq:1.36})  one obtains the recursions for the susy transformed fields as
\csep 
\begin{eqnarray}
\label{eq:1.42}
\ca{R}\delta\Phi^I&=&i\theta^{\alpha A}\Gamma^I_{A\D{A}}\delta\Psi_{\alpha\D{A}}+\Lambda\cdot\Phi^I
                              -\epsilon^{\alpha A}\ca{C}^I_{\alpha A}\ ,\nonumber\\
\ca{R}\delta\Psi_{\beta\D{A}}&=&\theta^{\alpha A}\Gamma^I_{A\D{A}}\nabla_{\alpha\beta}\delta\Phi^I
                     +\Lambda\cdot\Psi_{\beta \D{A}} -\epsilon^{\alpha A}\ca{C}_{\alpha\beta A\D{A}}\ .
\end{eqnarray}
Therefore, modulo super gauge transformations with the parameter $\Lambda$ of (\ref{eq:1.33}) the recursion relations are susy covariant in case that the matter constraints (\ref{c2}), (\ref{IC}) and the Bianchi identities 
(\ref{eq:1.27})  are satisfied.

The rest of the proof that the component e.o.m.~(\ref{eq:1.38}) imply superfield e.o.m.\ and superfield constraints, proceeds exactly as in the previous discussion of section \ref{sec:1.1} by simply replacing all derivative operators by covariant derivatives. 
Thus again, the superfield constraints are completely 
equivalent to the component multiplet with the e.o.m.~(\ref{eq:1.38}). 
We will see in the next section how deformations of the constraint 
(\ref{eq:1.25}) will modify these results and introduce non-trivial interactions.

\section{Interacting theories}
\label{sec2}

\subsection{Vector superfield with a modified constraint}
%
\label{subsec:modified}

In this section, we consider the vector superfields $\ca{A}_{\alpha A}$, $\ca{A}_{\alpha\beta}$ for 
which the constraint 
(\ref{eq:1.25}) is modified to
\begin{equation}
\label{defW}
 \{\nabla_{\alpha A} , \nabla_{\beta B} \} =
     2i \left(\delta_{AB} \nabla_{\alpha\beta} +  \vep_{\alpha\beta} W_{AB} \right) \ ,
\end{equation}
where $W_{AB}=-W_{BA}$ is an antisymmetric $SO(8)$-tensor.\footnote{This corresponds to a deformation of the 
free CS constraint (\ref{eq:1.25}) by the $({\bf 1}, {\bf 28})$ part of (\ref{convent}). 
In principle, one may also consider a deformation of the constraint by the $({\bf 3},{\bf 35})$ part of (\ref{convent})
and we come back to this possibility in the conclusions in section~\ref{sec4}.
For the main part of the paper we stay with the ansatz (\ref{defW}) as it turns out that 
the models we are interested in (BLG and SYM theory) precisely fit into this class.
}
There are two different situations in which the system (\ref{defW})
may appear. {First}, if $W_{AB}$ is a given function of the matter superfields of the theory, 
i.e.\ $W_{AB}=W_{AB}(\textstyle{\Phi^I,\Psi_{\alpha \D{A}}})$, the system (\ref{defW}) describes 
a {\em deformation} of the 
original constraint (\ref{eq:1.25}) which will in particular induce a (non-linear) deformation of the 
original (super)field equations of motion (\ref{eq:1.38}) by terms containing 
$W_{AB}$ and its (super-)derivatives.
This is the scenario we will be dealing with in this paper.
As we will see, as soon as the matter superfields are coupled to the 
gauge superfields, $W_{AB}$ is necessarily a function of them. 
In this case we will refer to 
the $SO(8)$-tensor $W_{AB}$ as the 
\emph{deformation potential}.

Alternatively, one might consider the vector multiplet independently and  
regard $W_{AB}$ as an independent field 
defined by equation (\ref{defW}), in which case this equation rather amounts to parametrizing a {\em weakening}
of the original constraint (\ref{eq:1.25}) to
\bea
\{\nabla_{\alpha A} , \nabla_{\beta B} \}\,
 \Big|_{({\bf 3},{\bf 35_s})} &=& 0\;.
\label{weakconstraint}
\eea
In that case, the dynamics induced
by (\ref{weakconstraint})  can be considered independently of the matter sector and will in particular 
lead to a different number of degrees of freedom
contained in the vector superfield. 

In either case the Bianchi identities impose conditions on $W_{AB}$ for the 
the constraint (\ref{defW}) being self consistent.

{\subsubsection*{Bianchi identities}} 
As in the free theory, the immediate nontrivial conditions on the 
superfields are given by the first two Bianchi identities in (\ref{eq:BI}), where (\ref{defW}) appears. 
Using the constraint (\ref{defW}) the first Bianchi identity imposes the condition  
\begin{eqnarray}
\label{WF}
\delta_{AB} \ca{F}_{\alpha\beta,\gamma C}
+\delta_{CA} \ca{F}_{\gamma\alpha,\beta B}  
&+&\delta_{BC} \ca{F}_{\beta\gamma,\alpha A}   =\nonumber\\
&&\vep_{\beta\gamma} \nabla\!_{\alpha A} W_{BC}+
\vep_{\gamma\alpha} \nabla\!_{{\beta B}} W_{CA}+
\vep_{\alpha\beta}\nabla\!_{\gamma C} W_{AB}\  .
\end{eqnarray}
Decomposing the terms of this equation analogously to (\ref{eq:1.4-2})
according to their $SO(8)$ representation content,
one deduces that solvability requires the ${\bf 160_s}$ to vanish within the 
the tensor product 
$\nabla_{\alpha A} W_{BC}\sim {\bf 8_s} \otimes {\bf 28} = {\bf 8_s}\oplus {\bf 56_s}\oplus {\bf 160_s}$.
This implies the existence of superfields 
$\lambda_{\alpha A}$, in the ${\bf{8_s}}$, and $\rho_{\alpha ABC}=\rho_{\alpha [ABC]}$,  in the ${\bf{56_s}}$, such 
that  the superderivative $\nabla_{\alpha A} W_{BC}$ satisfies the condition\footnote{Symmetrization and antisymmetrization
of indices is indicated by brackets $(\ )$ and $[\ ]$, respectively,  and is defined with total weight one,
i.e.\ $x_{(\alpha\beta)}=\frac12(x_{\alpha\beta}+x_{\beta\alpha})$, etc..}
\begin{equation}
\label{constraint160}
\nabla_{\alpha A} W_{BC}\;\Big|_{\bf 160_s} = 0\qquad \Longrightarrow\qquad 
\nabla_{\alpha A} W_{BC}  =  \delta_{A[B} \lambda_{C]\alpha}+\rho_{\alpha ABC} \ .
\end{equation}

This constraint will play a central role in the following.
In particular, if we consider $W_{AB}$ as a function of the matter fields of the theory,
this composite superfield must satisfy (\ref{constraint160}) in order for the system (\ref{defW}) to be consistent.
The Bianchi identity (\ref{WF}) then fixes the fermionic field strength  $\ca{F}_{\alpha\beta,\gamma A}$ to 
\begin{equation}
\label{FL}
    \ca{F}_{\alpha\beta,\gamma A} =-\vep_{\gamma(\alpha}\lambda_{\beta)A}\ .
\end{equation}
Using the constraint (\ref{defW}), the second Bianchi identity in (\ref{eq:BI}) writes as 
\begin{equation}
\label{bi2}
\nabla_{\alpha A}\ca{F}_{\gamma\delta,\beta B} +  \nabla_{\beta B}\ca{F}_{\gamma\delta,\alpha A}=
   2i(\delta_{AB}\ca{F}_{\gamma\delta,\alpha\beta}+\vep_{\alpha\beta}\nabla_{\gamma\delta}W_{AB})\ ,
\end{equation}
and with (\ref{FL}) implies the existence of another superfield $V_{AB}=V_{[AB]}$ in the ${\bf{28}}$,
such that 
\begin{equation}
\label{Dlambdarho}
\nabla_{\alpha A}\lambda_{\beta B} =
i(\delta_{AB} \ca{F}_{\alpha\beta}
+2\nabla_{\alpha\beta} W_{AB}
+\vep_{\alpha\beta}V_{AB}) \ .
\end{equation}
Here, $\ca{F}_{\alpha\beta}=\ca{F}_{(\alpha\beta)}$ 
denotes the vector dual to the bosonic field strength, i.e.\ 
$\ca{F}_{\alpha\beta}:=\vep^{\gamma\delta} \ca{F}_{\alpha\gamma, \beta\delta}$.
This duality is characteristic for three dimensions
and we will use this relation frequently in the following.

The first Bianchi identity identifies $\ca{F}_{\alpha\beta,\gamma A}$ 
with a single field (\ref{FL}) and thus, 
contrary to the free case (\ref{eq:1.27}), also the third Bianchi identity in (\ref{eq:BI}) 
gives a nontrivial condition on the superfields:
\begin{equation}
\label{eq:2.10}
\nabla_{\alpha A} \ca{F}_{\beta\gamma} =
\nabla_{\alpha(\beta} \lambda_{\gamma)A}+
\vep_{\alpha(\beta}\, \nabla_{\gamma)\delta} \lambda_A^{\delta}  \ .
\end{equation}

The equations (\ref{constraint160}), (\ref{FL}),  (\ref{Dlambdarho}) and (\ref{eq:2.10})  
are the consistency 
conditions for the constraint (\ref{defW}), which are imposed by the Bianchi identities. 

{\bf{Deformed super-CS e.o.m.}}
In the case that a deformation potential $W_{AB}=W_{AB}(\textstyle{\Phi^I,\Psi_{\alpha \D{A}}})$ 
is chosen  the derived superfields 
$\lambda_{\alpha A}$, $\rho_{\alpha ABC}$, etc.\ are also given functions of the matter superfields. 
In particular defines (\ref{Dlambdarho}) the super 
field strength $\ca{F}_{\alpha\beta}$ in terms of the matter superfields in the following form:
\begin{equation}
\label{eq:2.9}
\ca{E}_{\alpha\beta}:= \ca{F}_{\alpha\beta}+\fr{i}{8}\ \nabla_{(\alpha}^A\lambda_{\beta)A}= 
     \ca{F}_{\alpha\beta}-  \fr{i}{28}\ \nabla^A_{(\alpha}\nabla^B_{\beta)}W_{AB} =0\ ,
\end{equation}
where we used (\ref{constraint160}) to express $\lambda_{\alpha A}$ 
in terms of the deformation potential $W_{AB}$. 
As in the free case (\ref{eq:1.27}) one obtains the superfield e.o.m.\ 
in the gauge sector from the second  Bianchi 
identity and (\ref{eq:2.9}) explicitly shows, how the dynamics of the free Chern-Simons gauge field
is deformed by the presence of the deformation potential~$W_{AB}$.

A priori, with (\ref{eq:2.9}) the fourth Bianchi identity in (\ref{eq:BI}), which takes the form
\begin{equation}
\label{eq:2.11}
\nabla^{\alpha\beta}\ca{F}_{\alpha\beta}=0\ ,
\end{equation}
may give rise to yet another condition.
However, one can evaluate the l.h.s.\ of (\ref{eq:2.11}) using the constraint (\ref{defW}) and the 
conditions (\ref{eq:2.10}), (\ref{FL}) to show that (\ref{eq:2.11}) is identically fulfilled and does not impose 
additional conditions.

\subsubsection*{Integrability conditions}  

The integrability conditions 
of the constraints derived  from the  Bianchi identities, in particular (\ref{constraint160}) and (\ref{Dlambdarho}), 
determine the superderivatives of the various additional superfields and 
eventually allow to  define a closed recursive system for a systematic superfield 
expansion analogous to the procedure in section \ref{sec1}.
In the case that the gauge sector with the constraint (\ref{defW}) is considered as an independent system
these are genuine conditions on these superfields which correspond to independent degrees of freedom.
We give a thorough account on this scenario in 
appendix \ref{ap:WC}.

By contrast, in  choosing a certain deformation potential $W_{AB}(\textstyle{\Phi^I,\Psi_{\alpha \D{A}}})$  
satisfying the conditions  
(\ref{constraint160}),(\ref{Dlambdarho}) and (\ref{eq:2.10}), the ``sources'' on the r.h.s 
are derived from $W_{AB}$ and the integrability conditions are identically satisfied and give identities 
rather than conditions. In addition, the constraints   
(\ref{defW}) and (\ref{FL}) define  $\ca{R}\ca{A}_{\alpha A}$ and  $\ca{R}\ca{A}_{\alpha\beta}$ 
in terms  of the matter superfields and thus form together with
$\ca{R}\Phi^I$, $\ca{R}\Psi_{\alpha \D{A}}$  a closed recursive system.  We will carry out the detailed
analysis of the superfield expansion, component equations and the equivalence thereof to the constraints
in the next subsection, where we study the coupling between the gauge and matter sector.
We develop here the system of 
integrability conditions till the point we will need it for 
a  general discussion of the possible couplings to the matter sector.
Especially we want to clarify here which 
of the restrictions (\ref{constraint160}),(\ref{Dlambdarho}), (\ref{eq:2.10}) on the choice for the deformation potential  
$W_{AB}(\textstyle{\Phi^I,\Psi_{\alpha \D{A}}})$ are independent. 

The integrability condition of (\ref{Dlambdarho}) gives $\nabla_{\alpha A}V_{BC}$ and reproduces the 
third Bianchi identity (\ref{eq:2.10}). Analyzing  the 
integrability conditions of 
(\ref{constraint160}) determines $\nabla_{\alpha A}\rho_{\beta BCD}$ and 
reproduces the second Bianchi identity (\ref{Dlambdarho}) with $\ca{F}_{\alpha\beta}$ as given by 
the CS-e.o.m.~(\ref{eq:2.9}).  Consequently, the only remaining restriction on the 
choice of $W_{AB}(\textstyle{\Phi^I,\Psi_{\alpha \D{A}}})$ is the condition (\ref{constraint160}). 

The resulting covariant super derivatives of the various fields are:
\csep
\begin{eqnarray}
\label{icNA}
\nabla_{\alpha A}\, \rho_{\beta BCD} &=&
     3i\nabla_{\alpha\beta} W_{[BC}\delta_{D]A}
         -\ft{3i}2\vep_{\alpha\beta}\delta_{A[B}V_{CD]}
            +3i\vep_{\alpha\beta} \left[W_{A[B},W_{CD]}\right]+iU_{\alpha\beta\, ABCD}\ ,\nonumber\\[.5ex] 
\nabla_{\alpha A} V_{BC}  &=&  2\vep^{\beta\gamma} \nabla_{\alpha\beta}\left(
                              \delta_{A[B}\lambda_{C]\gamma}-\rho_{\gamma ABC}\right)
                                       -\left[ W_{BC} ,\lambda_{A\alpha}\right]
                                       - 4 \left[W_{A[B}, \lambda_{C]\alpha}\right]\ ,\nonumber\\[.5ex]
    \nabla_{\alpha A} U_{\beta\gamma\, BCDE}&=&
                          8\delta^{A[B} \nabla_{\alpha(\beta}\rho_{\gamma)}^{CDE]}
                           -4\delta^{A[B} \nabla_{\beta\gamma}\rho_{\alpha}^{CDE]}
                             +\tau_{\alpha\beta\gamma\, ABCDE} \nonumber\\
                                      &&+ 4\vep_{\alpha(\beta}\left(
                                      \ft43 [W^{A[B}, \rho_{\gamma)}^{CDE]}]
                                      - [W^{[BC} , \rho_{\gamma)}^{DE]A}]  
                                  +3\delta^{A[B} [W^{CD},\lambda_{\gamma)}^{E]}]\right)\ ,
\end{eqnarray}
where the last equation for the superfield $U_{\beta\gamma BCDE}=U_{(\beta\gamma) [BCDE]}$ has been 
obtained from the integrability 
condition for the $ \nabla_{\alpha A}\, \rho_{\beta BCD} $ equation. At this point superderivatives of the fields 
are determined up to 
the tensor $\tau_{\alpha\beta\gamma\, ABCDE}=\tau_{(\alpha\beta\gamma)\, [ABCDE]}$. This is all we need 
for a general discussion of the matter 
couplings and we refer 
to  appendix \ref{ap:WC} to see how the system closes.

We have thus shown, that deforming the free constraint (\ref{eq:1.25}) by choosing $W_{AB}$ to be a certain function 
$W_{AB}(\textstyle{\Phi^I,\Psi_{\alpha \D{A}}})$ of the matter superfields, the Bianchi identities 
are satisfied provided that $W_{AB}$ satisfies the constraint (\ref{constraint160}). 
The super field strengths  
are given by (\ref{FL}) and the deformed super Chern-Simons 
equations~(\ref{eq:2.9}). Consequently, the constraint (\ref{constraint160}) 
is the only condition on the choice of $W_{AB}$ 
for the deformation (\ref{defW}) to be self-consistent.

\subsection{Matter superfields and gauge matter coupling}
\label{sec:msf}

In this section we study the consequences of the deformation (\ref{defW}) for the 
matter sector and give a detailed discussion parallel to the sections \ref{sec:fcs} and \ref{sec:mc} of the coupled 
system  regarding component field equations, supersymmetry transformations and the equivalence thereof to the 
combined constraint 
system. As for the gauge sector the deformation will modify the dynamics by terms polynomial in the deformation potential 
$W_{AB}$ and its (super-)derivatives. Compatibility of the system will require $W_{AB}$ to satisfy additional 
algebraic constraints. 

\subsubsection*{Superfield constraints}
The most conceivable starting point for the matter sector
is to keep the covariantized constraint (\ref{c2}) for the scalar 
superfield $\Phi^{I}$
\begin{equation}
\label{dc2}
\nabla_{\alpha A} \Phi^I = i  \Gamma^{I}_{A\dot{A}}\,\Psi_{\alpha\D{A}}\ ,
\end{equation}
and deduce the consequences due to  the new vector superfield constraint (\ref{defW}). For a given 
constraint in the gauge sector, 
(\ref{dc2}) to a large extent determines the resulting dynamics of the system.

Using the gauge field constraint (\ref{defW}) the integrability condition of (\ref{dc2}) is now modified to
\begin{equation}
\label{dic}
2\, \delta_{AB} \nabla_{\alpha\beta}\Phi^I  + 2\, \vep_{\alpha\beta} \,W_{AB} \cdot \Phi^I
= \Gamma^{I}_{B\dot{A}} \nabla_{\alpha A} \Psi^{\dot{A}}_\beta
+\Gamma^{I}_{A\dot{A}} \nabla_{\beta B} \Psi^{\dot{A}}_\alpha \ .
\end{equation}
Repeating the analysis of section \ref{sec1} determines $\nabla_{\alpha A}\Psi_{\beta \D{A}}$ but also 
gives restrictions on 
the new (second) term on the l.h.s. Since the $\bf{160_v}$ in
\begin{equation}
W_{AB}\cdot\Phi^I \sim \bf{28}\otimes\bf{8_v}=\bf{8_v}\oplus\bf{56_v}\oplus\bf{160_v}\ ,
\end{equation}
is unpaired in equation (\ref{dic})
it has to vanish separately. 
In the following it will be often convenient to write $W_{AB}$ in the vector notation 
$W_{IJ}=\fr{1}{4}\Gamma^{IJ}_{AB}W_{AB}$ (see appendix~\ref{ap:gam}), such that 
the constraint on $W_{AB}\cdot\Phi_K$ writes as
\setlength\arraycolsep{0pt}
\begin{eqnarray}
\label{ac160}
&&W_{IJ}\cdot\Phi_K \Big|_{{\bf 160_v}} = 0 \qquad \Longrightarrow \nonumber\\  
&&\mathbb{P}_{160}^{[IJK]}(W_{IJ}\cdot\Phi_K): = 
W_{IJ}\cdot\Phi_K - W_{K[I}\cdot\Phi_{J]} +\fr{3}{7}\ \delta_{K[I}W_{J]L}\cdot\Phi^L\ =0\ .
\end{eqnarray}
In addition to the constraint (\ref{constraint160}) this will be the main restriction on the possible 
choices for the  deformation potential
$W_{AB}(\Phi^I,\Psi_{\alpha\D{A}})$, which fixes the details of the dynamics. In the following we will
refer to these two constraints (\ref{constraint160}), (\ref{ac160}), which determine the set of possible 
models, as the \emph{$W$-constraints}.  
The algebraic $W$-constraint  (\ref{ac160}) also shows that as soon as the matter sector is coupled 
to the gauge sector, 
the modification 
$W_{AB}$ of the gauge field constraint (\ref{defW}) has to be considered as a function of the the matter superfields 
which at least 
depends  on $\Phi^I$.

After some $SO(8)$-$\Gamma$-matrix algebra the integrability condition (\ref{dic}) yields
\begin{equation}
\label{dIC}
\nabla_{\alpha A}\Psi_{\beta\D{A}}=\Gamma^I_{A\D{A}}\nabla_{\alpha\beta}\Phi^I
  +\hal\vep_{\alpha\beta}\left(\fr{1}{7}\, \Gamma^I_{A\D{A}}\delta^{JK}+\fr{1}{6}\, \Gamma^{IJK}_{A\D{A}}\right)
        W_{IJ}\cdot\Phi_K
        \;,
\end{equation}
for the superderivative of the fermionic superfield.
Using the gauge field constraint (\ref{defW}) to express $x$-space covariant derivatives through 
covariant superderivatives 
and the various constraint relations and Bianchi identities of this section, one obtains  
the superfield equations for $\Psi_{\alpha\D{A}}$ and $\Phi^I$:
\csep
\begin{eqnarray}
\ca{E}_{\alpha\D{A}}:=&&\vep^{\beta\gamma}\nabla_{\alpha\beta}\Psi_{\gamma\D{A}}\nonumber\\
            && {}+\fr{3}{14}\, W_{\D{A}\D{B}}\cdot\Psi_{\alpha\D{B}} 
                + \fr{3 i}{16}\, \Gamma^I_{A\D{A}}\, \lambda_{\alpha A}\cdot \Phi^I
             + \fr{i}{336}\, \Gamma^{ABC}_{I\D{A}}\rho_{\alpha ABC}\cdot\Phi^I = 0\ \ ,\nonumber\\[2.5ex]
\ca{E}^I:=&&\nabla^2 \Phi^I 
                 - \fr{1}{8}\left( 3\, \Gamma^I_{A\D{A}}\ \lambda_{\alpha A}\cdot\Psi^\alpha_{\D{A}}
+ \fr{1}{21}\, \Gamma^{ABC}_{I\D{A}}\rho_{\alpha ABC}\cdot\Psi^\alpha_{\D{B}}\right) 
 \nonumber\\
                &&{} + \fr{3}{14}\, V^{IJ}\cdot\Phi^J - \fr{2}{49}\, W_{IJ}\cdot(W_{JK}\cdot\Phi^K) 
                        -  \fr{1}{28}\, W_{JK}\cdot(W_{JK}\cdot\Phi^I) = 0\  ,\nonumber\\
                        \label{dsfeom}
\end{eqnarray}
where  
$V^{IJ}:=\fr{1}{4}\, \Gamma^{IJ}_{AB}\, V_{AB}$, 
$W_{\D{A}\D{B}}:=\fr{1}{4}\, \Gamma^{IJ}_{\D{A}\D{B}}W_{IJ}$, are special cases of 
$SO(8)$ triality relations. In the same spirit we have defined the symbol 
$\Gamma^{ABC}_{I\D{A}}:=\Gamma^{IJ}_{[AB}\Gamma^J_{C]}{}_{\dot{A}}$,
see appendix \ref{ap:gam} for more details and 
several $\Gamma$-matrix identities which were employed in this calculation. Using the algebraic $W$-constraint 
(\ref{ac160}) one can recast  the scalar self-interaction involving $W_{IJ}$ in different forms. Equations 
(\ref{dsfeom}) together with (\ref{eq:2.9}) constitute the complete set of superfield e.o.m.

{\subsubsection*{Superfield expansion}} 
We again impose the 
``transverse'' gauge (\ref{eq:1.28}) to construct the superfield expansion via a recursive system. Contracting the constraints
for the matter fields (\ref{dc2}), (\ref{dIC}) and the gauge field constraint (\ref{defW}) with $\theta^{\alpha A}$, and  the 
Bianchi identity (\ref{FL}) with $\theta^{\gamma A}$, one obtains the recursion relations for the  superfields
$\Phi^I$, $\Psi_{\alpha\D{A}}$, $\ca{A}_{\alpha A}$ and $\ca{A}_{\alpha\beta}$:
\csep
\begin{eqnarray}
\label{eq:drec}
\ca{R}\Phi^I&=&i \theta^{\alpha A} \Gamma^I_{A\D{A}}\Psi_{\alpha\D{A}}\ \ ,\nonumber\\
\ca{R}\Psi_{\beta\D{A}}&=&\theta^{\alpha A}\Gamma^I_{A\D{A}}\nabla_{\alpha\beta}\Phi^I + 
\hal\, \theta^{\alpha A}\vep_{\alpha\beta}
     \left( \fr{1}{7}\, \Gamma^I_{A\D{A}}\delta^{JK}+\fr{1}{6}\, \Gamma^{IJK}_{A\D{A}}\right)
             W_{IJ}\cdot\Phi_{K}\ \ ,\nonumber\\
(1+\ca{R})\,\ca{A}_{\beta B}&=&2i (\theta^{\alpha}_{B}\ \ca{A}_{\alpha\beta}
          + \theta^{\alpha A}\vep_{\alpha\beta}W_{AB})\ \ , \nonumber\\
\ca{R}\ca{A}_{\alpha\beta}&=&\theta^{\gamma A}\vep_{\gamma (\alpha}\lambda_{\beta)A}\ \ ,
\end{eqnarray}
which generalize the recursions of the free theory (\ref{eq:1.29}) and (\ref{eq:1.36}).
The composite superfields of the gauge sector, such as $\lambda_{\alpha A}$, $\rho_{\alpha ABC}$, $V_{AB}$, etc.,  are 
now given functions of the matter superfields via the deformation potential $W_{AB}(\Phi^I,\Psi_{\alpha\D{A}})$,
\begin{eqnarray}
\label{eq:2.2.1}
&&\lambda_{\alpha A}=\fr{2}{7}\ \nabla_{\alpha B} W_{BA}\ , \quad 
          \rho_{\alpha ABC}=\nabla_{\alpha [A}W_{BC]}\ , \nonumber\\ 
&& V_{AB}= - \fr{i}{2}\ \vep^{\alpha\beta}\nabla_{\alpha A}\nabla_{\beta C}W_{CB} \ ,\ \ \mathrm{etc.},
\end{eqnarray}
as can be seen from equations (\ref{constraint160}), (\ref{Dlambdarho}) and (\ref{icNA}). The 
recursion relations for
these composite superfields as well as for $W_{AB}$ and $\ca{F}_{\alpha\beta}$ are determined 
by the recursions of the fundamental 
superfields (\ref{eq:drec}), but on the  
constraint surface they are equivalently given by the contraction 
of (\ref{constraint160}), (\ref{Dlambdarho}), (\ref{eq:2.10}) 
and  (\ref{icNA}) with $\theta^{\alpha A}$. Off the constraint surface, and thus when deriving the 
constraints from the component field equations, this is no longer true as we will see.

To second order in $\theta$, the superfield expansion can be expressed in terms of the 
composite fields explicitly given in (\ref{eq:2.2.1}):
\csep
\begin{eqnarray}
\label{dsfexp}
\Phi^I&=&\phi^I+i \theta^{\alpha A}\Gamma^I_{A\D{A}} \psi_{\alpha \D{A}}
+\fr{i}{2}\ \theta^{\alpha A}\theta^{\beta B}\ \Gamma^{IJ}_{AB}\  \ov{\nabla}_{\alpha\beta}\phi^J
\nonumber\\ &&\ \  -\fr{i}{4}\, \theta^{\alpha A}\theta^{\beta B}\vep_{\alpha\beta}
     \Big( \fr{1}{7}\ \delta_{AB}\  \ov{W}_{IJ}\cdot\phi^J-\fr{1}{6}\ \Gamma^{ILMN}_{AB}\ \ov{W}_{LM}\cdot\phi_{N}\Big)
+\ldots\;, \nonumber\\
\Psi_{\beta\D{A}}&=&\psi_{\beta\D{A}}+\theta^{\alpha A}\Gamma^I_{A\D{A}}\ov{\nabla}_{\alpha\beta}\phi^I
\nonumber\\
&&\ \ +\hal\, \theta^{\alpha A}\vep_{\alpha\beta}
( \fr{1}{7}\, \Gamma^I_{A\D{A}}\delta^{JK}+\fr{1}{6}\, \Gamma^{IJK}_{A\D{A}})\ \ov{W}_{IJ}\cdot\phi_K
+\ldots \;,\nonumber\\
\ca{A}_{\alpha\beta}&=& A_{\alpha\beta}
  +\theta^{\gamma C}\vep_{\gamma(\alpha}\ov{\lambda}_{\beta)C} \nonumber\\
&& +\fr{i}{2}\, \theta^{\gamma C}\theta^{\delta D}\,
\Big(\hal \vep_{\gamma\delta}\ \delta_{CD} F_{\alpha\beta}
-2\vep_{\gamma(\alpha}\ov{\nabla}_{\beta)\delta}\ov{W}_{CD}
    +\vep_{\gamma(\alpha}\vep_{\beta)\delta}\ov{V}_{CD}\Big)+\ldots\;,
\end{eqnarray}
while for $\ca{A}_{\beta B}$, the expansion of is formally the same as in abelian case (\ref{AaA}) of appendix~\ref{ap:WC},  see 
also  (\ref{eq:psiexp}) for more details. 
To obtain explicit expressions one has to compute the lowest order components of the composite 
fields in (\ref{eq:2.2.1}). 
To this end we assume here and in the following that $W_{AB}$ depends on $\Phi^I$ only and not on the fermionic
superfield $\Psi_{\beta\D{A}}$, i.e.\
\begin{equation}
\label{wf}
W_{AB}=W_{AB}\ (\Phi^I)\ . 
\end{equation}
The explicit cases that we are going to study in this work  fall into this class of 
deformation potentials $W_{AB}$. Using
(\ref{dc2}) and 
with $\del_{Ia}:=\del/\del\phi^{Ia}$, where the index $a$ refers to the representation of the gauge (structure) group, 
the projection on the lowest components for the composite fields takes the form
\csep
\begin{eqnarray}
\label{2.2.2}
&&\ov{W}_{AB}= W_{AB}(\phi) \ ,\ \
\ov{\lambda}_{\alpha B}=\fr{2i}{7}\, \psi_{\alpha\D{A}}^a\Gamma^I_{A\D{A}}\ \del_{Ia}\ov{W}_{AB} \ , \ \
\ov{\rho}_{\alpha ABC}= i\psi_{\alpha\D{A}}^a\ \del_{Ia}\ov{W}_{[BC}\Gamma^I_{A]\D{A}}\ ,\nonumber\\
&&\ov{V}_{AB}=[\ov{W}_{AC},\ov{W}_{CB}]
 - \vep^{\alpha\beta} \psi_{\alpha\D{A}}^a\psi_{\beta\D{B}}^b\ \Gamma^I_{A\D{A}}\Gamma^J_{C\D{B}}
    \ \del_{Ia}\del_{Jb}\ov{W}_{CB}\nonumber\\
&&\quad\quad\quad\quad+\fr{i}{7}\ (\ov{W}_{IJ}\cdot\phi^J)^a\ \del_{Ia}\ov{W}_{AB}
   -\fr{i}{6}\Gamma^{ILMN}_{AC}\ (\ov{W}_{LM}\cdot\phi_N)^a\ \del_{Ia}\ov{W}_{BC}\ .
\end{eqnarray}

With the above relations and the superfield expansions (\ref{dsfexp}) one obtains from (\ref{dsfeom}) 
in a straightforward way the 
component field e.o.m.\ for the component fields $\phi^I$ and $\psi_{\alpha \D{A}}$. 
The CS-e.o.m.\ is the lowest component of (\ref{eq:2.9}) and can
be computed analogously to (\ref{2.2.2}). Together, the full system of component e.o.m.\ is given by
\csep
\begin{eqnarray}
\label{eq:1}
&&\ov{\ca{E}}_{\alpha\beta} =
    F_{\alpha\beta}+\frac{1}{28}\left(\Gamma^{IJ}_{AB}\ov{\nabla}_{\alpha\beta} \phi^{Ia} \del_{Ja}\ov{W}_{AB}
-i \psi^a_{\D{A}(\alpha}\psi^b_{\beta)\D{B}}\ \Gamma^I_{A\D{A}}\Gamma^J_{B\D{B}}\ 
             \del_{Ia}\del_{Jb}\ov{W}_{AB}\right)=0\ ,
\nonumber\\
&&\ov{\ca{E}}_{\alpha\D{A}}=0\ ,\quad\quad \ov{\cal{E}}{}^I=0\ \ .
\end{eqnarray}

The supersymmetry transformations for the independent component fields $\phi^I$, $\psi_{\alpha\D{A}}$ and $A_{\alpha\beta}$ 
are again obtained
from the superfield expansion, (\ref{dsfexp}), by acting with $\epsilon^{\alpha A}Q_{\alpha A}$ and modding out 
a restoring super gauge transformation with gauge parameter
\begin{equation}
\label{eq:2}
\Lambda=\epsilon^{\alpha A}\ca{A}_{\alpha A}=
 i\epsilon^{\alpha A}\theta^{\beta B}(\delta_{AB}A_{\alpha\beta}+\vep_{\alpha\beta}\ov{W}_{AB})+\ldots\ \ ,
\end{equation}
which is formally the same as for the free CS-multiplet (\ref{eq:1.33}) but has a more non-trivial superfield expansion,
see (\ref{AaA}). The obtained component supersymmetry transformations are,
\csep
\begin{eqnarray}
\label{eq:3}
\delta\phi^I&=&i \epsilon^{\alpha A}\ \Gamma^I_{A\D{A}}\psi_{\alpha\D{A}}\ ,\nonumber\\
\delta\psi_{\beta \D{A}}&=&\epsilon^{\alpha A}\Big(\Gamma^I_{A\D{A}}\ov{\nabla}_{\alpha\beta}\phi^I
      +\hal\ \vep_{\alpha\beta}\left(\fr{1}{7}\, \Gamma^I_{A\D{A}}\delta^{JK}
     +\fr{1}{6}\, \Gamma^{IJK}_{A\D{A}}\right)\ov{W}_{IJ}\cdot\phi_K\Big)\ ,\nonumber\\
\delta A_{\alpha\beta}&=&\fr{2i}{7}\, \epsilon^{\gamma B}
            \vep_{\gamma(\alpha}\psi^a\!\!{}_{\beta) \D{A}}\ \Gamma^I_{A\D{A}}\ \del_{Ia}\ov{W}_{AB} \ \ ,
\end{eqnarray}
and again resemble the recursion relations of the associated superfields (\ref{eq:drec}). 
Equations (\ref{eq:1}) and (\ref{eq:3}) show how the deformation potential $W_{AB}$ modifies the 
dynamics and supersymmetry transformations
of the component fields compared to the  minimally coupled free CS-multiplet (\ref{eq:1.38}) and (\ref{eq:1.40}).
The deformation potential 
$\ov{W}_{AB}=W_{AB}(\phi^I)$
cannot be chosen arbitrarily but inherits the lowest components  of the $W$-constraints (\ref{constraint160}) 
and (\ref{ac160}).
These conditions are also necessary for the component field equations (\ref{eq:1}) to be invariant under the 
supersymmetry transformations (\ref{eq:3}). The algebraic $W$-constraints (\ref{ac160}) is the same for the lowest component 
fields, since
it just constrains the functional form of $W_{AB}(\phi^I)$. The lowest component of the differential $W$-constraint 
(\ref{constraint160}) is straightforwardly
obtained by using (\ref{dc2}). Together, one finds for $\ov{W}_{AB}=W_{AB}(\phi^I)$ the conditions
\begin{equation}
\label{eq:4}
\mathbb{P}_{160}^{[IJK]}(\ov{W}_{IJ}\cdot\phi_K)=0\  ,\qquad 
 \mathbb{P}_{160}^{[ABC]}(\Gamma^I_{A\D{A}}\del_{Ia}\ov{W}_{BC})=0\ \ ,
\end{equation}
where the projector $\mathbb{P}_{160}^{[RST]}$, acting on three indices $R,S,T$
referring to the same representation, was introduced in (\ref{ac160}).

\subsubsection*{Equivalence to component e.o.m.}

In this part we prove that the component multiplet $(\phi^I,\psi_{\alpha\D{A}},A_{\alpha\beta})$ satisfying the 
e.o.m.~(\ref{eq:1}) with the conditions (\ref{eq:4}) for the deformation potential, 
and the supersymmetry transformations (\ref{eq:3}) is equivalent to our constraint system,
in particular the gauge field constraint (\ref{defW}) and the matter field constraint (\ref{dc2}) and consequently 
their Bianchi
identities and integrability conditions. The reader who is only interested  in the mere fact of this equivalence may skip the 
details of the proof presented here.

As in the previous sections we construct superfields 
$\ca{A}_{\alpha A}$, $\ca{A}_{\alpha\beta}$, $\Phi^I$ and $\Psi_{\alpha\D{A}}$ out of the 
component multiplet according to 
the recursion relations\footnote{We do not intend to carry this out explicitly but
use the recursions (\ref{eq:drec}) as an implicit definition of the superfields. The explicit calculation
would be rather messy, especially since off the constraint surface one cannot use the previously given recursion 
relations for the composite fields, as we will demonstrate now.}  (\ref{eq:drec}). 
One can ask again if this definition of
superfields is susy covariant and mutatis mutandis one obtains the result analogous to (\ref{eq:1.32}) and (\ref{eq:1.42})  
that these superfields are susy covariant modulo supergauge transformations with the parameter (\ref{eq:2}) 
if the constraints (\ref{defW}), (\ref{FL}), (\ref{dc2}) and (\ref{dIC}) are satisfied. 

To demonstrate the equivalence between component field equations and the constraints we again 
construct a recursive system for the constraints- and superfield e.o.m.\ expressions.
Due to the non-trivial coupling of the gauge and matter sector, and in particular due to the conditions on the 
deformation potential 
$W_{AB}$, the situation is quite involved and we introduce a 
more symbolic notation such that the structure of the system remains clear. From the gauge sector the following
expressions, resembling (\ref{defW}), (\ref{FL}), (\ref{Dlambdarho}) and (\ref{constraint160}) will occur in 
the recursive system
\csep
\begin{eqnarray}
\label{eq:5}
G^{(1)} = G_{\alpha A,\beta B} &:=& \{\nabla_{\alpha A},\nabla_{\beta B}\} 
- 2i(\delta_{AB}\nabla_{\alpha\beta} + \vep_{\alpha\beta}W_{AB})\  , \nonumber\\
G^{(2)} = G_{\alpha\beta,\gamma A} &:=& \ca{F}_{\alpha\beta,\gamma A} 
+ \vep_{\gamma(\alpha}\lambda_{\beta)A} \ , \nonumber\\
\ca{E}^{\mathrm{cs}} = \ca{E}_{\alpha\beta} &:=& \ca{F}_{\alpha\beta} - X_{\alpha\beta} \ , \nonumber\\
G= G_{\alpha A,BC} &:=& \nabla_{\alpha A}W_{BC}-(\delta_{A[B}\lambda_{C]\alpha}+\rho_{\alpha ABC})\ ,
\end{eqnarray}
where we have introduced the abbreviation $X_{\alpha\beta}=-\fr{i}{8}\, \nabla^C\!\!\!{}_{(\alpha}\lambda_{\beta)C}$, and the 
other composite fields were given in (\ref{eq:2.2.1}). The expressions
of the matter sector, resembling (\ref{dc2}), (\ref{dIC}), (\ref{dsfeom}) and (\ref{ac160}), are:
\csep
\begin{eqnarray}
\label{eq:6}
C^{(1)} = C_{\alpha A}^I &:=& \nabla_{\alpha A}\Phi^I-i\Gamma^I_{A\D{A}}\Psi_{\alpha\D{A}} \ ,\nonumber\\
C^{(2)} = C_{\alpha A,\beta\D{A}} &:=& \nabla_{\alpha A}\Psi_{\beta\D{A}} 
               - \Gamma^I_{A\D{A}}\nabla_{\alpha\beta}\Phi^I \nonumber\\
  &&\hspace{2cm} - \hal\vep_{\alpha\beta}\left(\fr{1}{7}\ \Gamma^I_{A\D{A}}\delta^{JK}
                +\fr{1}{6}\ \Gamma^{IJK}_{A\D{A}}\right)
        W_{IJ}\cdot\Phi_K  \ ,\nonumber\\
 \ca{E}^{\mathrm{ferm}}&:=& \ca{E}_{\alpha\D{A}}\ , \qquad  \ca{E}^{\mathrm{bos}}:= \ca{E}^I \ ,\nonumber\\
C_{IJK}&:=&\mathbb{P}_{160}^{[IJK]}(W_{IJ}\cdot\Phi_K) \ .
\end{eqnarray}
The explicit expressions for $\ca{E}_{\alpha\D{A}}$, $ \ca{E}^I$ were given in (\ref{dsfeom}). In the following,
the 
detailed index structure of the occurring expressions will not be important and in general 
we stick to the notation on the
l.h.s of these definitions.

To determine the action of the recursion operator
$\ca{R}$ (\ref{eq:1.28}) on the expressions (\ref{eq:5}), (\ref{eq:6}) we will need  the superderivatives of 
the composite fields off the constraint surface, i.e.\ the analogs of (\ref{Dlambdarho}), (\ref{eq:2.10}) and (\ref{icNA}), but 
with $\ca{F}_{\alpha\beta}$ replaced with $X_{\alpha\beta}$\footnote{\label{fn}
Note that on the constraint surface means also 
$\ca{F}_{\alpha\beta}=X_{\alpha\beta}$ with $X_{\alpha\beta}$ given below (\ref{eq:5}), i.e.\ the CS-e.o.m., see also the 
discussion above (\ref{icNA}).}. These equations were obtained as 
consecutive integrability conditions of the differential $W$-constraint (\ref{constraint160}). 
Off the constraint surface one has to start instead from 
$G$ in (\ref{eq:5}). Keeping track also of the other constraints one finds the following modifications of
(\ref{Dlambdarho}), (\ref{eq:2.10}), (\ref{icNA}):
\csep
\begin{eqnarray}
\label{eq:10}
\nabla_{\alpha A}\lambda_{\beta B}&\rightarrow&  \nabla_{\alpha A}\lambda_{\beta B}+\{G^{(1)}  W+\nabla G\}\ \ ,
\nonumber\\
\nabla_{\alpha A}\rho_{\beta BCD}&\rightarrow& \nabla_{\alpha A}\rho_{\beta BCD} + \{G^{(1)}  W+\nabla G\}\ \ ,
\nonumber\\
\nabla_{\alpha A}X_{\beta\gamma}&\rightarrow&\nabla_{\alpha A}X_{\beta\gamma}
+\{G^{(1)} \lambda+\nabla(G^{(1)}  W+\nabla G)+G^{(2)}  W\}\ \ ,\nonumber\\
\nabla_{\alpha A}V_{BC}&\rightarrow&\nabla_{\alpha A}V_{BC}
+\{G^{(1)} \lambda+\nabla(G^{(1)}  W+\nabla G)+G^{(2)}  W\}\ \ ,\nonumber\\
\nabla_{\alpha A}U_{\beta\gamma BCDE}&\rightarrow&\nabla_{\alpha A}U_{\beta\gamma BCDE}
+\{G^{(1)} \lambda+\nabla(G^{(1)}  W+\nabla G)+G^{(2)}  W + G  W\}\ , \nonumber\\
\end{eqnarray}
where $\nabla$ symbolically stands for a superderivative $\nabla_{\alpha A}$ with unspecified indices.
In addition we need the expression for $\ca{R}\ca{F}_{\alpha\beta}$. 
As a consequence of the recursive definitions of the independent superfields (\ref{eq:drec})
certain contractions of the constraints with $\theta^{\alpha A}$  vanish identically, i.e.\ 
$\theta^{\alpha A}C^I_{\alpha A}=\theta^{\alpha A}C_{\alpha A,\beta\D{A}}=\theta^{\alpha A}G_{\alpha A,\beta B}=
\theta^{\delta D}G_{\alpha\beta,\delta D}=0$. With this one  finds by acting 
with $\nabla_{\gamma\delta}$ on $\ca{R}\ca{A}_{\alpha\beta}$ in (\ref{eq:drec}) 
\begin{equation}
\label{eq:8}
\ca{R}\ca{F}_{\alpha\beta} =  \theta^{\delta D}(\nabla_{\delta(\alpha} \lambda_{\beta)D}+
\vep_{\delta(\alpha}\, \nabla_{\beta)\gamma} \lambda_D^{\gamma})\ \ ,
\end{equation}
which is the same as on the constraint surface, i.e.\ the 
equation obtained by contraction of  (\ref{eq:2.10}) with $\theta^{\alpha A}$.

For the $W$-constraints the results follow directly from the conditions (\ref{eq:4}) 
on the  lowest components of the deformation potential $W_{AB}$.
The algebraic $W$-constraint in (\ref{eq:6}) is identically zero, i.e.\  $C_{IJK}=0$, due to the first condition in (\ref{eq:4}). 
The last equation  in (\ref{eq:5}) can be read as 
$G=\mathbb{P}_{160}^{[ABC]}(\nabla_{\alpha A}W_{BC})$ and thus as a consequence of the 
second condition in (\ref{eq:4}) the differential  $W$-constraint takes the form
\begin{equation}
\label{eq:9}
G_{\alpha A,BC}=\mathbb{P}_{160}^{[ABC]}(C_{\alpha A}^{Ia}\del_{Ia}W_{BC})\quad 
                      \mathrm{or}\quad G\sim C^{(1)}\del W\ ,
\end{equation}
where the second expression is of the symbolic form that we will use in this section.

We now have all the ingredients needed to compute the action of $\ca{R}$ on the 
other
expressions (\ref{eq:5}) and (\ref{eq:6}).
Using the recursions (\ref{eq:drec}) and the relations derived in this part, 
one finds for the gauge sector
\csep
\begin{eqnarray}
\label{eq:11}
(2+\ca{R})\ G^{(1)}&\sim&\theta \{G^{(2)}+G\}\ \ ,\nonumber\\
(1+\ca{R})\ G^{(2)}&\sim& \theta \{\ca{E}^{\mathrm{cs}}+(G^{(1)}  W+\nabla G)\}\ \ ,\nonumber\\
 \ca{R}\ \ca{E}^{\mathrm{cs}}&\sim&\theta \{G^{(1)} \lambda
      +G^{(2)}  W+\nabla(G^{(1)}  W+\nabla G)\}\ ,
\end{eqnarray}
which are obtained more or less straightforwardly. For the matter sector one obtains
\csep
\begin{eqnarray}
\label{eq:12}
(1+\ca{R})\ C^{(1)}&\sim&\theta  C^{(2)}\ \ ,\nonumber\\
(1+\ca{R})\ C^{(2)}&\sim& \theta \{\ca{E}^{\mathrm{ferm}}+\nabla_{\alpha\beta}C^{(1)}+G^{(2)} \Phi
      +(G \Phi+W  C^{(1)})\}\ \ ,\nonumber\\
 \ca{R}\ \ca{E}^{\mathrm{ferm}}&\sim&  \theta \{\ca{E}^{\mathrm{cs}} \Phi
    +\ca{E}^{\mathrm{bos}}+\nabla(G \Phi+W  C^{(1)})+(G^{(1)}  W \Phi
       +W  C^{(2)})\}\ ,\nonumber\\
   \ca{R}\ \ca{E}^{\mathrm{bos}}&\sim& \theta \{ \nabla_{\alpha\beta}\ca{E}^{\mathrm{ferm}}
      +\ca{E}^{\mathrm{cs}} \Psi+\nabla\nabla(G \Phi+W  C^{(1)})\nonumber\\
       &&+\nabla(G^{(1)}  W \Phi+W  C^{(2)})
        +(G^{(2)}  W \Phi+G  W \Phi+W  W  C^{(1)})\}\  ,\nonumber\\
\end{eqnarray}
where $\nabla_{\alpha\beta}$ symbolically stands for a bosonic covariant derivative, the given indices have no 
specific meaning. The first relation in (\ref{eq:12}) is straightforwardly obtained and uses the fact that $C_{IJK}=0$, as explained 
above (\ref{eq:9}). The derivation of the  other relations is rather involved and uses, in this order, the first, second and 
third superderivative of the just mentioned relation, i.e.\ $\nabla C_{IJK}=0$,  $\nabla\nabla C_{IJK}=0$ and 
$\nabla\nabla\nabla C_{IJK}=0$. Via (\ref{eq:10}) these produce  a number of constraints which we extracted here, 
the remaining  terms are found to cancel with the help of an algebraic computation 
using Mathematica.

The notation  used in (\ref{eq:11}) and (\ref{eq:12})  is rather formal, the suppressed index structure appears in all kind of 
combinations. This is 
enough information to show recursively that the whole system of constraints (\ref{eq:5}), (\ref{eq:6}) vanishes to all orders in 
$\theta$ as a consequence of the equations for the component fields (\ref{eq:1}), (\ref{eq:4}). In the first step 
one sees that to lowest order all expressions in (\ref{eq:5}) and (\ref{eq:6}) are zero due to  (\ref{eq:1}), (\ref{eq:4})
or the recursion relations  (\ref{eq:11}), (\ref{eq:12}):
\begin{equation}
\label{eq:13}
\ov{C}{}^{(1)}=\ov{G}= \ov{G}{}^{(1)}=\ov{C}{}^{(2)}= \ov{G}{}^{(2)}=\ov{\ca{E}}{}^{\mathrm{ferm}}
=\ov{\ca{E}}{}^{\mathrm{cs}}=\ov{\ca{E}}{}^{\mathrm{bos}} =0\ .
\end{equation}
In the sequence given here for 
the lowest component it is easy to show 
using  (\ref{eq:11}), (\ref{eq:12}) and (\ref{eq:9}), that 
to order $(n+1)$ in $\theta$ all expressions in  (\ref{eq:5}), (\ref{eq:6}) are zero if they vanish at 
order $n$  (the only subtlety one has to be careful about is the appearance of the superderivatives 
$\nabla$ in   (\ref{eq:5}), (\ref{eq:6}), which brings in higher order coefficients). 
With (\ref{eq:13})  this inductively proves that all expressions in  (\ref{eq:5}), (\ref{eq:6}),
vanish to all orders in $\theta$ due to the component field equations  (\ref{eq:1}), (\ref{eq:4}), and thus shows
the equivalence of the component field formulation and the constraints (\ref{defW}) and (\ref{dc2}) and 
all their consequences. 
\\

Concluding this section, we have shown that the  weaker gauge field constraint (\ref{defW}) is consistent only if the 
deformation potential $W_{AB}$ satisfies the differential $W$-constraint  (\ref{constraint160}). Coupling to 
the matter system via the 
same constraint as in the free CS case, (\ref{dc2}), further imposes  the algebraic $W$-constraint (\ref{ac160}) on
$W_{AB}$  and thus necessarily makes the deformation potential a function of the matter superfields. This results in the 
interacting CS- and matter superfield e.o.m.~(\ref{eq:2.9}) and  (\ref{dsfeom}). 
For the case that $W_{AB}$ is a function exclusively of the scalar superfield $\Phi^I$ we gave the component field 
e.o.m.\ and the supersymmetry transformations (\ref{eq:1}), (\ref{eq:3}), (\ref{eq:4}) and demonstrated the equivalence
of the component field equations to the superfield constraints. The generalization of these considerations to a more general
deformation potential 
$W_{AB}$, depending also on the fermionic superfield $\Psi_{\alpha\D{A}}$ is straightforward. 
In the next section we will give explicit solutions
to the $W$-constraints  which will imply the conformal BLG-model and $\ca{N}=8$ SYM theory in its dual formulation, respectively.

\section{Solutions to the superspace constraints}\label{sec3}

In this section, we present different solutions
to the obtained superspace constraints and show how
all known examples of three-dimensional ${\cal N}=8$ gauge theories 
fit into our framework.
Let us start by reviewing the structure of superspace constraints 
identified so far. The matter sector of these three-dimensional gauge theories 
is described by a scalar superfield subject to the constraint (\ref{dc2})
\bea
\nabla_{\alpha A}\, \Phi^I \Big|_{{\bf 56_c}} &=& 0  \;.
\label{conP}
\eea
The full theory is then identified by specifying their gauge algebra $\mathfrak{g}$ (\ref{eq:1.16})
as a subalgebra of ${\mathfrak{gl}}(N,\mathbb{R})\oplus_s{\mathfrak{t}}(8N)$
and by choosing $W_{AB}(\Phi^I, \Psi_{\alpha\dot{A}})$ in (\ref{defW}) as a 
function of the matter superfields of the theory.
This choice of the deformation potential $W_{AB}$ is not arbitrary but 
must satisfy two independent superfield conditions, the $W$-constraints 
(\ref{constraint160}) and (\ref{ac160}):
\bea
\nabla_{\alpha A} W_{BC}\;\Big|_{\bf 160_s} &=& 0\;,
\label{conW1}\\[1ex]
W_{IJ} \cdot \Phi_K\;\Big|_{\bf 160_v} &=& 0
\;.
\label{conW2}
\eea
The first equation requires that the deformation potential $W_{AB}$ depends on the matter fields
such that (\ref{conW1}) is satisfied as a consequence of (\ref{conP}).
In contrast, equation (\ref{conW2}) also explicitly contains the action of the
gauge group on the matter fields and will thus put further restrictions on
the possible gauge groups.
We will see in explicit examples, that the conditions (\ref{conW1}), (\ref{conW2})
are truly independent as there are solutions to either one of them that do not solve the other equation.

\subsection{Conformal gauge theories}

In this section we consider gauge groups $G$ that are subgroups of $GL(N,\mathbb{R})$,
$N$~being the number of scalar super-multiplets,
such that the superfield ${\cal A}_{\alpha A}$ can be represented as a matrix
in the adjoint representation of~$G$.
Accordingly, we label by indices $a, b, \dots$, the representation of $G$
in which the matter superfields $\Phi_I$ and $\Psi_{\alpha\dot{A}}$ transform. 
Matter and gauge superfields are thus denoted as
$\Phi_I^{a},\, \Psi_{\alpha\dot{A}}^a$, and 
$({\cal A}_{\alpha A})^a{}_b,\; ({\cal A}_{\alpha\beta})^a{}_b$,
respectively.

The constraint (\ref{defW}) implies that the composite field $W_{AB}$ has canonical dimension one. Given that 
the scalar fields have canonical dimension one half in three dimensions, scale invariance implies that 
with a polynomial ansatz  $W_{AB}$ is
bilinear in the scalar superfields $\Phi_I^{a}$, 
with the most general ansatz
given by
\bea
(W_{IJ})^a{}_b &\equiv& f^a{}_{b,cd}\,\Phi^{c}_I \Phi^{d}_J
\; , 
\label{Wconf}
\eea
where the dimensionless constants $f^a{}_{b,cd}$ have to be antisymmetric in the last two indices, i.e.\ 
$ f^a{}_{b,[cd]}= f^a{}_{b,cd}$.
Gauge covariance requires that $f^a{}_{b,cd}$ is an invariant tensor of the
gauge group $G$, and per construction $W_{IJ}$ has to be an element of the Lie algebra and therefore 
$f^a{}_{b,dc}\in{\mathfrak{g}}$ for any $d$ and $c$.
Together, this translates into a quadratic condition for the tensor 
$f^a{}_{b,cd}$
\bea
f^g{}_{c,ab}f^e{}_{f,gd}-f^g{}_{d,ab}f^e{}_{f,gc}&=&
f^g{}_{f,cd}f^e{}_{g,ab} - f^g{}_{f,ab}f^e{}_{g,cd}
\;,
\label{quadf}
\eea
which can be obtained by explicitly evaluating the action of $W_{IJ}$ on a $W_{KL}$
and comparing this to the adjoint action by commutator. The same relation was obtained in 
\cite{Bergshoeff:2008cz} for the embedding tensor in a component field approach.  

It is straightforward to check, that (\ref{Wconf}) is a solution to (\ref{conW1})
as a consequence of (\ref{conP}): as $\nabla_{\alpha A} W_{BC}$ is composed of
a single $\Phi^I$ and a single $\Psi_{\alpha\dot{A}}$, w.r.t.\ $SO(8)$ it transforms 
in the tensor product 
${\bf 8_v}\otimes {\bf 8_c}={\bf 8_s}+{\bf 56_s}$, which does not contain a ${\bf 160_s}$\,.
To solve the remaining constraint (\ref{conW2}) we evaluate the action of (\ref{Wconf})
on a scalar field
\bea
(W_{IJ} \cdot \Phi_K)^a &=& f^a{}_{b,cd}\,\Phi^b_I\Phi^c_J\Phi^d_K\;.
\eea
This shows that the tensor $f^a{}_{b,cd}$ needs to satisfy complete antisymmetry
in the last three indices $f^a{}_{b,cd}=f^a{}_{[bcd]}$, such that
\bea
(W_{IJ} \cdot \Phi_K)^a &=&
f^a{}_{[bcd]}\,\Phi^b_{[I}\Phi^c_J\Phi^d_{K]}
\;,
\eea
transforms in the ${\bf 8_v}^{\otimes_{\rm alt} 3}= {\bf 56_v}$ of $SO(8)$,
thus satisfying (\ref{conW2}).
For such a tensor  $f^a{}_{b,cd}$, the quadratic equation (\ref{quadf}) reduces to the so-called {\em fundamental identity}.
The same condition on a  tensor $f^a{}_{bcd}$, interpreted as a structure constants of a three-algebra,  
has been used  in~\cite{Gran:2008vi} in a component formulation of the equations of motion.
This shows how the constructions of~\cite{Bagger:2007jr,Gustavsson:2007vu,Gran:2008vi} are embedded into our 
superspace analysis.
The existence of an action furthermore requires the existence of a metric $h_{ab}$ and total antisymmetry of the tensor
$f_{abcd}\equiv h_{ae} f^e{}_{[bcd]}$\,. It has been shown in a number of papers
(see e.g.~\cite{Papadopoulos:2008sk,Gauntlett:2008uf}),
that for a positive definite metric $h_{ab}$, equation (\ref{quadf}) admits no other solutions than the
compact $SO(4)$ of the original construction of~\cite{Bagger:2007jr,Gustavsson:2007vu}.
Solutions of (\ref{quadf}) with indefinite metric have been found and studied in
\cite{Gomis:2008uv,Benvenuti:2008bt,Ho:2008ei}.

In order to complete the construction of this example, we evaluate the general formulae of the last section
for the particular choice (\ref{Wconf}). From (\ref{eq:2.2.1}),
we obtain
\bea
(\lambda_{\alpha A})^a{}_b &=& if^a{}_{bcd}\,\Gamma^I_{A\dot{A}}\,
\Psi_{\alpha \dot{A}}^c \Phi_I^d\;,\qquad
(\rho_{\alpha\,ABC})^a{}_b ~=~ -\ft12 i  f^a{}_{bcd}\,\Gamma_{I\dot{A}}^{ABC} 
\,\Psi_{\alpha \dot{A}}^c \Phi_I^d
\;,
\nonumber\\[.5ex]
(V_{AB})^a{}_b &=& 
-\ft12 i f^a{}_{bcd}\, \vep^{\alpha\beta} \Gamma_{AB}^{\dot{A}\dot{B}}\,
\Psi^c_{\alpha\dot{A}}\Psi^d_{\beta\dot{B}}
+
\ft1{4} f^a{}_{bcd}f^c{}_{efg}\,
\Gamma^{IJ}_{AB} \,  \Phi_I^e \Phi_J^f \Phi_K^g \Phi_K^d
\;,
\eea
as well as the first order Chern-Simons equations of motion (\ref{eq:2.9})
\bea
({\cal F}_{\alpha\beta})^a{}_b &=& - f^a{}_{bcd} \Big(\Phi_I^c \,\nabla_{\alpha\beta} \Phi_{I}^d
+i \Psi^c_{\alpha\dot{A}}\Psi^d_{\beta\dot{A}} \Big)
\;.
\eea
After some calculation, the bosonic equations of motion (\ref{dsfeom}) take the form
\bea
\label{eomBL}
\nabla^2\, \Phi^a_I &=& 
\fr{i}{2}\,   \vep^{\alpha\beta}
\Gamma^{IJ}_{\dot{A}\dot{B}}\,
f^a{}_{bcd}\,
\Psi_{\alpha \dot{A}}^b  \Psi_{\beta \dot{B}}^c \Phi_J^d
+\ft{1}{4}\,f^a{}_{bcd}f^b{}_{efg}\,
\Phi_J^c \Phi_J^f  \Phi_K^d \Phi_K^g \Phi_I^e
\;,
\eea
and coincide with the result of \cite{Gran:2008vi}.  For the theories with action, 
they exhibit the Yukawa couplings and the sextic scalar potential of~\cite{Bagger:2007jr}.

\subsection{Yang-Mills gauge theories}\label{sec:YM}

It has been shown in \cite{deWit:2003ja,Nicolai:2003bp} that three-dimensional Yang-Mills gauge theories have an
equivalent formulation as matter-coupled Chern-Simons gauge theories with non-semisimple
gauge group 
\bea
G = G_{\rm YM}\ltimes \mathbb{T}_k
\;,
\label{GCS}
\eea
where $\mathbb{T}_k$ denotes a set of
$k\equiv {\dim}\,G_{\rm YM}$ translations, transforming in the adjoint representation of $G_{\rm YM}$\,.
This allows to embed also Yang-Mills gauge theories into the general superspace formulation
presented above. 
In the context of M2 branes, this duality has been discussed in \cite{Ho:2008ei,Ezhuthachan:2008ch}.

In order to realize (\ref{GCS}) as a subgroup of $GL(N,\mathbb{R})\ltimes \mathbb{T}(8N)$,
we start from matter fields $\Phi_I^a$, $\Psi_{\alpha\dot{A}}^a$ in the adjoint representation (thus $N=k$), 
with the index $a$ now labelling the adjoint
representation of the Yang-Mills gauge group $G_{\rm YM}$,
and $f_{ab}{}^c$ denoting the Yang-Mills structure constants. 
To obtain the subalgebra $\mathfrak{t}$  
associated with the subgroup $\mathbb{T}_k\subset\mathbb{T}(8k)$
we choose a fixed $SO(8)$-vector $\xi_I$ and define the  generators ${\cal T}_a$ of $\mathfrak{t}$ as
\begin{equation}
\label{eq:subal}
\ca{T}_a=\xi_I \ca{T}^I_a\ \ ,
\end{equation}
with a constant vector $\xi_I$,
and where the $ \ca{T}^I_a$  span the full Lie algebra of $\mathbb{T}(8k)$.
The  gauge superfields in the covariant derivatives as defined in (\ref{eq:1.15}) are thus chosen to be
\begin{eqnarray}
\label{eq:3.15}
\ca{A}_{\alpha A}=\ca{A}^M_{\alpha A}iT_M&=&\ca{A}^a_{\alpha A}\ i T_a
             +\ca{B}^{a}_{\alpha A}\ i{\cal T}_{a}
                           =: \hat{\ca{A}}_{\alpha A}+\ca{B}_{\alpha A}\;,
\end{eqnarray}
with the Yang-Mills and the translation generators acting on the scalar superfield as
\bea
T_{a} \cdot \Phi_I^b &=& f_{ac}{}^b \Phi_I^c\;,
\qquad
{\cal T}_{a} \cdot \Phi_I^b = \xi_I \delta_a^b
\;,
\label{actionTT}
\eea
respectively. The constant vector $\xi_I$ breaks $SO(8)$ down to $SO(7)$.
The algebra of the generators ({\ref{eq:1.16}}) hence splits into the semidirect sum as
\begin{equation}
\label{eq:3.17}
[T_M,T_N]=f_{MN}^{\ \ \ \ K} T_K
\quad \leftrightarrow\quad
\left\{ 
                \begin{array}{rl} 
                          &[T_a,T_b]= f_{ab}{}^c \,T_c\\ 
                          &[T_a,{\cal T}_{b}]=f_{ab}{}^c\, {\cal T}_c\\
                          & [{\cal T}_a,{\cal T}_b]=0
                \end{array} \right. \;.
\end{equation}
The bosonic gauge superfield $\ca{A}_{\alpha\beta}=\ca{A}^M_{\alpha \beta}T_M$ 
is decomposed analogously to (\ref{eq:3.15}), except for the  factor of $i$ (\ref{eq:1.15}). 
With regard to the separation of the gauge superfields we can write the covariant derivatives  accordingly,
\begin{equation}
\label{eq:3.19}
\nabla_{\alpha A}=\hat{\nabla}_{\alpha A}+\ca{B}_{\alpha A}\ ,\quad
     \nabla_{\alpha \beta}=\hat{\nabla}_{\alpha \beta}+\ca{B}_{\alpha \beta}\ ,
\end{equation}
where $\hat{\nabla}_{\alpha A}$ contains only $\hat{\ca{A}}_{\alpha A}$, etc.. The action on the 
superfield $\Phi_I$ then takes the form
\begin{equation}
\label{eq:3.20}
\nabla_{\alpha A}\Phi^a_I=\hat{\nabla}_{\alpha A} \Phi_I^a+i \xi_I \ca{B}^a_{\alpha A}\ ,
\end{equation}
and accordingly for the bosonic superfield connection $ \nabla_{\alpha \beta}$. 
On all other fields, which are 
neutral under shifts generated by the ${\cal T}_{k}$, the action of $\nabla_{\alpha A}$ 
and $\hat{\nabla}_{\alpha A}$ coincides.
The explicit form of the gauge transformations (\ref{eq:1.21}) is then given by
\begin{eqnarray}
\label{eq:3.21}
&&\delta\Phi^I=\Lambda\cdot\Phi^I+\xi^I C\ ,\quad
            \delta\Psi_{\alpha\D{A}} = \Lambda\cdot\Psi_{\alpha\D{A}}\,,\nonumber\\
&&\delta\hat{\ca{A}}_{\alpha A} = -\hat{\nabla}_{\alpha A}\cdot\Lambda\,, \qquad
\delta \ca{B}_{\alpha A} = i\hat{\nabla}_{\alpha A}C+\Lambda\cdot \ca{B}_{\alpha A} \, ,
\end{eqnarray}
and analogous transformations for 
the bosonic superfields
$\hat{\ca{A}}_{\alpha \beta}$, ${\ca{B}}_{\alpha \beta}$. 
These transformations lead to a homogeneous covariant transformation
of the super covariant derivatives of $\Phi^I$, i.e.\
\begin{equation}
\label{eq:3.22}
\delta(\nabla_{\alpha A}\Phi^I)=\Lambda\cdot(\nabla_{\alpha A}\Phi^I)\ ,
\end{equation}
which thus is neutral under local shifts in $\Phi^I$.
As for the covariant derivatives, also the conventional field strengths acquire extra terms 
only in the case of their action on the scalar superfields $\Phi^I$. 
With the definitions (\ref{eq:3.15}), (\ref{eq:3.17}) one obtains for the anti-commutator
\setlength\arraycolsep{1pt}
\begin{eqnarray}
\label{eq:3.23}
\{\nabla_{\alpha A},\nabla_{\beta B}\}\cdot\Phi^I         
 &=& \left(2i\delta_{AB}\del_{\alpha\beta} +D_{\alpha A}\ca{A}_{\beta B}
   +D_{\beta B}\ca{A}_{\alpha A}+\{\ca{A}_{\alpha A},\ca{A}_{\beta B}\}\right) \cdot\Phi^I 
   \nonumber\\[.5ex]
&\equiv&         2i\delta_{AB}\nabla_{\alpha\beta}\Phi^I + \hat{\cal{F}}_{\alpha A, \beta B}\cdot\Phi^I+\xi^I {\cal H}_{\alpha A, \beta B}
     \ ,
\end{eqnarray}
with the split of field strength into ${\cal{F}}_{\alpha A, \beta B}=\hat{\cal{F}}^a_{\alpha A, \beta B}T_a
+{\cal H}^a_{\alpha A, \beta B}{\cal T}_a$, i.e.\
\bea
\hat{\cal{F}}_{\alpha A, \beta B} &=&
D_{\alpha A}\hat{\ca{A}}_{\beta B}
   +D_{\beta B}\hat{\ca{A}}_{\alpha A}+\{\hat{\ca{A}}_{\alpha A},\hat{\ca{A}}_{\beta B}\}
   -2i\delta_{AB} \hat{\cal {A}}_{\alpha\beta}
   \;,
\nonumber\\
{\cal H}_{\alpha A, \beta B} &=&
\hat{\nabla}_{\alpha A}\ca{B}_{\beta B}+\hat{\nabla}_{\beta B}\ca{B}_{\alpha A}
-2i\delta_{AB} {\cal {B}}_{\alpha\beta}
\;.
\label{FH}
\eea
Similarly, we split the bosonic field strength ${\cal F}_{\alpha \beta, \gamma\delta}$ into a part 
$\hat{{\cal F}}_{\alpha \beta, \gamma\delta}$ corresponding to the standard non-abelian Yang-Mills field strength
of the gauge field $\hat{\ca{A}}_{\alpha\beta}$ and 
${\cal H}_{\alpha \beta, \gamma\delta}=
\hat{\nabla}_{\alpha \beta}\ca{B}_{\gamma\delta}-\hat{\nabla}_{\gamma\delta}\ca{B}_{\alpha\beta}$ such that
\begin{eqnarray}
\label{eq:3.24}
{\cal F}_{\alpha \beta, \gamma\delta}\cdot \Phi^I
&=&\hat{{\cal F}}_{\alpha \beta, \gamma\delta}\cdot \Phi^I+\xi^I {\cal H}_{\alpha \beta, \gamma\delta}\,.
\end{eqnarray}

It remains to find a solution for the tensor $W_{IJ}$ living in the algebra (\ref{eq:3.17})
that satisfies the constraints  (\ref{conW1}) and (\ref{conW2}). 
Our proposal is the following
\bea
W_{IJ} &=& W_{IJ}^{(0)}{}^a \, T_a + W_{IJ}^{(1)}{}^a\, {\cal T}_a
\equiv
-2 \Phi^{a [I}  \xi^{J]}\,T_a + f_{bc}{}^{a} \Phi^{I b} \Phi^{J c}  \, {\cal T}_a
\;,
\label{WYM}
\eea
where the superscripts $^{(0),(1)}$ refer to the $G_{\rm YM}$-covariant grading of the algebra (\ref{eq:3.17}).
The $SO(8)$ vector $\xi^I$ has canonical dimension $\fr{1}{2}$.
It is straightforward to verify that this function satisfies the constraint (\ref{conW1})
as a consequence of (\ref{conP}). The argument is as in the last section: 
it follows with (\ref{conP}) that $\nabla_{\alpha A} W_{BC}$ w.r.t.\ $SO(8)$ transforms 
in the tensor product ${\bf 8_v}\otimes {\bf 8_c}={\bf 8_s}+{\bf 56_s}$, which does not contain a ${\bf 160_s}$\,.
Moreover, with (\ref{actionTT}) one checks that
\bea
(W_{IJ}\cdot \Phi_K)^a
&=&  
W_{IJ}^{(0)}{}^b f_{bc}{}^a \Phi_K^{c}
+ W_{IJ}^{(1)}{}^a\,\xi_K
\nonumber\\[1ex]
&=&3 f_{bc}{}^a \Phi_{[I}^{b}\Phi_{J}^{c} \xi^{\phantom{b}}_{K]}
=
(W_{[IJ}\cdot \Phi_{K]})^a
\;,
\eea
is completely antisymmetric in $[IJK]$, i.e.\ transforms in the ${\bf{56_v}}$, and thus also satisfies the 
constraint (\ref{conW2}).
This fixes the relative factor in (\ref{WYM}).
Finally, gauge covariance of the ansatz (\ref{WYM}) requires
\bea
f_{[ab}{}^d\,f_{c]d}{}^e &=& 0
\;,
\eea
the standard Jacobi identities for the structure constants of $G_{\rm YM}$.
To complete the construction, we evaluate the general formulae of section \ref{sec:msf}
for the solution (\ref{WYM}). From (\ref{eq:2.2.1}) we obtain
\bea
\lambda_{\alpha A} &=&
-i\Gamma^{A}_{I\dot{A}} \left(
\xi^{I}\Psi^{a}_{\alpha\dot{A}}  \,T_a 
-
i f_{bc}{}^{a} \,\Psi^{b}_{\alpha\dot{A}} \Phi^{I c}  \, {\cal T}_a \right)
\;,\nonumber\\[.5ex]
\rho_{\alpha\,ABC} &=& 
\ft12 i\Gamma^{ABC}_{I\dot{A}} \left(
\xi^{I} \Psi^{a}_{\alpha\dot{A}}  \,T_a
-i f_{bc}{}^{a} \, \Psi^{b}_{\alpha\dot{A}} \Phi^{I c}  \, {\cal T}_a \right)
\;,
\eea
and 
\bea
V_{AB}  &=&
-\ft3{4}  f_{bc}{}^a \Gamma^{KL}_{AB}\,
\xi^{I}\xi^{\phantom{a}}_{[I}   
\Phi_{K}^{b}\Phi_{L]}^{c}  
\,T_a
\nonumber\\
&&{}
+\left( \ft3{4} 
f_{bc}{}^{a} f_{de}{}^b
\Gamma^{KL}_{AB}\,
\xi^{\phantom{a}}_{[I}  \Phi_{K}^{d}\Phi_{L]}^{e} \Phi^{I\,c}   
-\ft12 i  \vep^{\alpha\beta} f_{bc}{}^{a} \,\Gamma^{AB}_{\dot{A}\dot{B}}\,
\Psi^{b}_{\alpha\dot{A}} \Psi_{\beta \dot{B}}^{c}  \right) {\cal T}_a  
\;.
\eea
The first order CS-equations of motion (\ref{eq:2.9}) yields
\bea
{\cal F}_{\alpha\beta} &=&
- 
\xi^{I} \, \nabla_{\alpha\beta}  \Phi^{I a}  \,T_a 
-
f_{bc}{}^{a} \left(\Phi^{I b} \nabla_{\alpha\beta}  \Phi^{I c}  
+i \Psi^{b}_{\alpha\dot{A}} \Psi_{\beta \dot{A}}^{c} \right)  {\cal T}_a 
\;.
\label{FfirstorderYM}
\eea
Finally, the bosonic equations of motion (\ref{dsfeom}) reduce to
\bea
\nabla^2\, \Phi_I^a &=& 
\fr{i}{2} \,\vep^{\alpha\beta}
\Gamma^{IJ}_{\dot{A}\dot{B}} \,
f_{bc}{}^a\,
\xi^{J}\Psi^{b}_{\alpha\dot{A}} 
\Psi^c_{\beta{\dot{B}}} 
+\fr{3}{2}\,
f_{bc}{}^d f_{de}{}^a 
\xi^{N}\xi^{\phantom{a}}_{[N}   
\Phi_{I}^{b}\Phi_{J]}^{c}  
\Phi_J^e
\;.
\label{eomYM}
\eea

In order to show the equivalence to the standard formulation of three-dimensional $\ca{N}=8$
Yang-Mills theories, one uses part of equations (\ref{FfirstorderYM}) to integrate out
the vector field ${\cal B}_{\alpha\beta}$. 
Explicitly, we split the $SO(8)$ index $I\rightarrow (i,8)$ with $i=1, \dots, 7$,
set $\xi^I=\gym\ \delta^{I8}$, with a dimension $\fr{1}{2}$ constant $\gym$, 
and fix the gauge freedom $\delta \Phi^I = \xi^I C$ by setting $\Phi^8=0$.
Note that this gauge differs from the ``transverse'' gauge (\ref{eq:1.28}), which imposes also  
$\theta^{\alpha A} {\cal B}_{\alpha A}=0$ and was used in the  previous 
analysis  to construct the superfield expansion.

Using (\ref{eq:3.20}) the $T_a$ component of equation (\ref{FfirstorderYM}) reduces 
in this new gauge to
\bea
\hat{\cal F}_{\alpha\beta} &=&
- \gym^{\ 2}{\cal B}_{\alpha\beta}
\;,
\eea
and can be used to eliminate the gauge field ${\cal B}_{\alpha\beta}$
from all equations. In particular, with ${\cal H}_{\alpha \beta, \gamma\delta}$ as given above 
(\ref{eq:3.24}) the remaining component of 
equation (\ref{FfirstorderYM})
takes the form
\bea
\fr{1}{ \gym^{\ 2}}\ \vep^{\gamma\delta}
\hat{\nabla}_{\gamma(\alpha}\hat{\cal F}_{\beta) \delta}
&=&
\fr{1}{2} [\Phi^{i} , \hat{\nabla}_{\alpha\beta}  \Phi^{i}]  
+  i \Psi_{(\alpha}^{\dot{A}} \Psi_{\beta)}^{\dot{A}} 
\;,
\eea
in which we recognize the standard second-order Yang-Mills equations of motion
for the remaining gauge field $\hat{\ca{A}}_{\alpha\beta}$\,. The scalar field equations of motion
are obtained from (\ref{eomYM}) after imposing $\xi^I \Phi_I=0$, and exhibit the quartic
potential in the scalar fields~$\Phi^i$.

It is instructive to study this redualization of the three-dimensional degrees of freedom
on a more fundamental level 
directly in terms of the superfield constraints. Upon setting $\Phi^8=0$,
the scalar constraint (\ref{conP}), or explicitly (\ref{dc2}), implies that
\bea
i\gym {\cal B}^a_{A\alpha} &=& \nabla_{A\alpha} \Phi^{8\,a} ~=~ i\Gamma^8_{A\dot{A}}\,\Psi^{\dot{A}\,a}_\alpha
\;,
\eea
i.e.\ the vector superfield $\ca{B}_{A\alpha}$ which gauges the translations is identified with the 
fermion superfield $\Psi^{\dot{A}}_\alpha$\,.
With (\ref{dIC}) we thus obtain from (\ref{FH})
\bea
{\cal H}_{\alpha A,\beta B} &=& 
{\hat{\nabla}}_{\alpha A} {\cal B}_{\beta B}+{\hat{\nabla}}_{\beta B} {\cal B}_{\alpha A} 
-2i \delta_{AB}{\cal B}_{\alpha\beta}
\nonumber\\
&=&
2i \delta_{AB}  (\gym^{-1}\nabla_{\alpha\beta} \Phi^{8\,a} -\ca{B}^a_{\alpha\beta})\,{\cal T}_a
+\gym^{-1}\fr{i}{6} \vep_{\alpha\beta} (\Gamma^{IJK}\Gamma^8)_{[AB]}\,
(W_{IJ}\cdot \Phi_K)^a\,{\cal T}_a
\nonumber\\
&=&
\gym^{-1} \fr{i}{2} \vep_{\alpha\beta} (\Gamma^{IJK}\Gamma^8)_{[AB]}\,
f_{bc}{}^a\, \Phi_{[I}^{b}\Phi_{J}^{c}\, \xi^{\phantom{b}}_{K]}\,{\cal T}_a
\nonumber\\
&=&
2i \,\vep_{\alpha\beta}
\,W_{AB}^{(1)}{}^a \,{\cal T}_a
\;.
\eea
I.e.\ the constraint (\ref{defW}) is automatically satisfied for the ${\cal T}_a$ component 
of the super-field strength. The remaining part of this superfield constraint yields
\bea
\hat{\cal F}_{A\alpha,B\beta} &=&
\ft12i \vep_{\alpha\beta}
\Gamma^{IJ}_{AB}\,W_{IJ}^{(0)}{}^a\,T_a
~=~ i \gym  \vep_{\alpha\beta}
\Gamma^{8i}_{AB}\,\Phi^{i}
\;,
\eea
or equivalently
\bea
\{ \hat\nabla_{A\alpha} , \hat\nabla_{B\beta} \} &=&
2i \delta_{AB}\hat\nabla_{\alpha\beta}
+ i \gym  \vep_{\alpha\beta} \Gamma^{8i}_{AB}\,\Phi^{i} 
\;.
\label{conYM}
\eea
If we take this equation as a definition for the scalar fields $\Phi^{i}$, 
the Bianchi identities for (\ref{conYM}) induce the matter superfield constraint (\ref{dc2}).
In this respect, equation (\ref{conYM}) may thus be considered as a weaker
version of the constraint (\ref{eq:1.25}), which accordingly gives rise to 
Yang-Mills dynamics rather than to a Chern-Simons dynamics for the gauge fields involved.
Moreover, we recognize in (\ref{conYM}) the remnant of the superfield constraint 
underlying ten-dimensional super Yang-Mills theory~\cite{Harnad:1985bc,Witten:1985nt}
\bea
\{ \nabla_{{\cal A}} , \nabla_{{\cal B}} \} &=&
2i \,\Gamma^{{\cal I}}_{{\cal AB}}\,\nabla_{{\cal I}}
\;,
\label{10dconstraint}
\eea
with $SO(9,1)$ vector and spinor indices {\footnotesize${\cal I}$} and {\footnotesize${\cal A}$}, respectively,
after breaking the Lorentz group $SO(9,1)\rightarrow SO(2,1)\times SO(7)$
and truncating the partial derivatives w.r.t.\ the seven internal coordinates.
The scalar fields $\Phi^{i}$ represent the seven internal components of the
ten-dimensional vector.

\section{Conclusions and outlook}\label{sec4}

In this paper, we have given a systematic analysis of the 
${\cal{N}}=8$ superspace constraints 
in three space-time dimensions.
The general coupling between vector and scalar supermultiplets is encoded
in the deformation potential $W_{\!AB}$ which is a function of the matter fields subject to 
the $W$-constraints (\ref{conW1}) and (\ref{conW2}).
The full equations of motion are given by equations
(\ref{eq:2.9}) and (\ref{dsfeom}).
We have given two solutions (\ref{Wconf}) and (\ref{WYM})
to these constraints corresponding
to  conformal matter Chern-Simons theories and to three-dimensional maximally
supersymmetric
Yang-Mills theory, respectively.
The presented results and the universal formalism in which all
known $\ca{N}=8$ three-dimensional gauge theories
have been embedded suggest a number of 
possible generalizations and directions of further research
of which we list a few in the following.

\begin{itemize}

\item
In the course of this paper we have met and analyzed various different constraints for the 
super field strength ${\cal F}_{\alpha A,  \beta B}$.
In its strongest version (\ref{eq:1.25}) the field strength ${\cal F}_{\alpha A, \beta B}$ is set to zero 
which gives rise to a (first order) Chern-Simons dynamics of the bosonic gauge field.
A weaker version of the constraint is (\ref{weakconstraint}) which allows for a non-vanishing part 
in the irreducible $({\bf 1,28})$. As shown in appendix~\ref{ap:WC}, this leads to an enlarged vector multiplet
with essentially no dynamics (apart from certain first order constraint equations
on the higher order components of the multiplet).
Yet another version of the constraint has been encountered in (\ref{conYM}) upon breaking 
$SO(8)$ down to $SO(7)$ and allowing an irreducible $({\bf 1,7})$ in the super field strength.
As discussed above, this is related to a ten-dimensional origin of the theory and 
induces a (second order) Yang-Mills dynamics for the bosonic gauge field.
In order of increasing constraints, these cases may be tabulated as
\bea
\{ \nabla_{A\alpha} , \nabla_{B\beta} \} \Big|_{(3,35)\phantom{+(1,28)}} = 0 &&
\qquad \Longrightarrow \quad
\mbox{no dynamics}\;,
\nonumber\\[.5ex]
\{ \nabla_{A\alpha} , \nabla_{B\beta} \} \Big|_{(3,35)+(1,21)} = 0 &&
\qquad \Longrightarrow \quad
\mbox{Yang-Mills dynamics}\;,
\nonumber\\[.5ex]
\{ \nabla_{A\alpha} , \nabla_{B\beta} \} \Big|_{(3,35)+(1,28)} = 0 && 
\qquad \Longrightarrow \quad
\mbox{free Chern-Simons dynamics}
\;,
\label{ccc}
\eea
and show how the field content and the dynamics becomes more 
restrictive as a function of the constraints.
It would be very interesting to perform a similar analysis for other versions of the constraint
upon breaking the original form under various subgroups of $SO(8)$ and to study the
resulting multiplet structures, their dynamics and a possible higher-dimensional origin.

\item
As shown in appendix~\ref{ap:WC}, the first constraint in (\ref{ccc}) admits the 
representation as a partial flatness condition for the integrability of an 
auxiliary linear system. 
No such representation is known for the constraint  (\ref{defW}) with $W_{AB}$ being a deformation potential,
as we have studied it in this paper. However, as we have discussed in section~\ref{sec:YM},
for the particular solution (\ref{WYM}) of the $W$-constraints, the super field strength may be 
brought into the form (\ref{conYM}) which descends from the 
zero-curvature condition on super null lines (\ref{10dconstraint})  of the linear system underlying the
ten-dimensional Yang-Mills equations of motion \cite{Witten:1985nt}. 
Dimensional reduction does not guarantee the existence of a linear auxiliary system and a corresponding twistor
space description. For example for the $\ca{N}=4$ SYM theory  in four dimensions
no such system is known, only the $\ca{N}=3$ superspace formulation has been described  in these geometric terms so far
\cite{Witten:1978xx}. However, the dimensional reduction of the the ten-dimensional SYM superspace constraints
to six dimensions, describing six-dimensional $\ca{N}=2$ SYM, can be reformulated as a linear 
auxiliary 
system\footnote{With $x^{ij}=x^{[ij]}=\hal\vep^{ijkl}x_{kl}$ being a six-dimensional vector ($i,j=1,\ldots,4$)
one finds  that the integrability conditions of  
$x^{ij}\nabla_{j\alpha}\ca{S}=x_{ij}\nabla^j_{\D{\alpha}}\ca{S}=x^{ij}\nabla_{ij}\ca{S}=0$ are equivalent to the 
superspace constraints for the six-dimensional $\ca{N}=2$ SYM theory as given in \cite{Harnad:1985bc}, 
\emph{iff} $x^{ij}$ is a null vector. 
The geometry of these null-vectors and the corresponding twistor space 
were discussed in \cite{Hughston:1986hb}, it is natural  to expect that 
there exists a twistor space formulation of the six-dimensional $\ca{N}=2$ SYM theory.}. 
In three dimensions a twistorial description of SYM has been given in $\ca{N}=6$ superspace \cite{Saemann:2005ji}.
However, it is an interesting question if there exists an auxiliary linear system and an associated twistor space 
description for the solution (\ref{Wconf}) of the $W$-constraints which eventually would give rise to a
linear system and associated twistor space formulation underlying the equations of motion of the conformal BLG model.
The covariance of our formalism suggest a study of this question analogous to SYM theories.

\item
In this paper we have studied the interactions between scalar and vector superfields
induced by a deformation (\ref{defW}) of the super field strength.
A natural generalization of this ansatz would also include the remaining irreducible term
\bea
\{\nabla_{\alpha A} , \nabla_{\beta B} \} &=&
2i \left(\delta_{AB} \nabla_{\alpha\beta} + 
\vep_{\alpha\beta} W_{AB}
+ J_{\alpha\beta\,AB}
\right)
\;,
\label{defWJ}
\eea
with a tensor $J_{\alpha\beta\,AB}=J_{(\alpha\beta)(AB)}$,
traceless in $(AB)$,
that is now likewise given as a function of the matter fields.
An analysis similar to the one performed in the main text, shows that 
in presence of a non-vanishing $J_{\alpha\beta\,AB}$ the differential $W$-constraint (\ref{conW1})
is modified to
\bea
\vep^{\beta\gamma}\,\nabla_{A\beta} J_{\gamma\alpha\,BC}\;\Big|_{\bf 160_s}
&=& 
\nabla_{\alpha A} W_{BC}\;\Big|_{\bf 160_s}
\;,
\nonumber\\
\nabla_{A(\alpha} J_{\beta\gamma)\,BC}\;\Big|_{\bf 112_s} &=& 0
\;,
\label{conWJ}
\eea
where the projectors on the l.h.s.\ refer to the irreducible 
parts of the tensor product
${\bf 8_s} \otimes {\bf 35_s} =
{\bf 8_s} \oplus {\bf 112_s} \oplus {\bf 160_s}$ 
in which $\nabla_{\alpha A} J_{\beta\gamma\,BC}$ transforms w.r.t.\ $SO(8)$\,.
Likewise, upon coupling to scalar superfields,
the algebraic $W$-constraint (\ref{conW2}) is extended to
\bea
W_{AB} \cdot \Phi^I\,\Big|_{{\bf 160_v}} &=& 0
~=~
J_{\alpha\beta\,AB} \cdot  \Phi^I\,\Big|_{{\bf 224_v}} \;.
\label{WJPhi}
\eea
We expect that similar to the analysis presented in the text, these constraints
will be sufficient to guarantee consistency of the system~(\ref{defWJ})
coupled to scalar superfields.
It remains an open question to find solutions of the extended set of constraints 
(\ref{conWJ}), (\ref{WJPhi}) that would give rise to more general ${\cal N}=8$ theories.

\item
Along similar lines, the system (\ref{conP})--(\ref{conW2}) can be generalized 
by deforming the matter superfield constraint (\ref{conP}), i.e.\ 
by allowing more general contributions
\bea
\nabla_{\alpha A}\, \Phi^I  &=& \Gamma^I_{A\dot{A}} \Psi_{\alpha \dot{A}}
+ \Gamma^{\dot{A}\dot{B}\dot{C}}_{I A}\,\Theta_{\alpha \dot{A}\dot{B}\dot{C}}
\;,
\label{modPP}
\eea
where now $\Theta_{\alpha \dot{A}\dot{B}\dot{C}}$ is considered as a function of
the superfields $\Phi^I$, $\Psi_{\alpha \dot{A}}$ 
(subject to a number of differential and algebraic constraints).
A similar strategy has been used in \cite{Cederwall:2001bt}
in order to constrain the higher order $\alpha'$ corrections
to ten-dimensional super Yang-Mills theory. 
In the present context, a viable strategy in order to describe
higher order corrections to the models may be to implement the algebraic
$W$-constraint (\ref{conW2}) by adequate choice of the deformation potential $W_{\!AB}$ 
while solving the differential $W$-constraint (\ref{conW1}) for this functional
by suitably tuning the $\Theta$ contribution in (\ref{modPP}) that
modifies (\ref{conP}). In this context it is also possible to consider 
non-polynomial generalizations of the ansatz (\ref{Wconf}) which are scale invariant. 
The verification of the conformal symmetry of the resulting models can be 
conveniently carried out by representing the superconformal algebra on the $\ca{N}=8$
superspace. These steps represent a possibility for determining quantum corrections without 
relying on perturbation theory.

\item
The generic scalar field equations of motion (\ref{dsfeom}) that we have
derived as a consequence of the superspace constraints exhibit various terms
containing the deformation potential $W_{\!AB}$, as well as the derived quantities $\lambda_{\alpha A}$,
$\rho_{\alpha ABC}$ and $V^{IJ}$. However, when explicitly evaluating these terms
for the explicit models in (\ref{eomBL}) and (\ref{eomYM}), we observe that all the
terms give rise to only two distinct contributions to the equations of motion,
a purely bosonic term and a single term bilinear in the fermions.
This raises the question if this reduction of the general equation
is related to some (yet undiscovered) 
underlying structure of the generic theory
or if there exist more general solutions to the $W$-constraints (\ref{conW1}), (\ref{conW2})
for which the different terms of (\ref{dsfeom}) do give contributions of different type.
The question may be related to the fact that both our explicit solutions
(\ref{Wconf}) and (\ref{WYM}) satisfy an algebraic equation which is actually stronger than
(\ref{conW2}) and reads
\bea
W_{IJ} \cdot \Phi_K\;\Big|_{{\bf 8_v}+{\bf 160_v}} &=& 0
\;.
\label{W8}
\eea
It would be highly interesting to understand if (\ref{W8}) is a (hidden)
consequence of the constraints (\ref{conW1}), (\ref{conW2}) or if the latter
admit solutions with a non-trivial component in the ${\bf 8_v}$. With regard to the supersymmetry 
transformations (\ref{eq:3}) this would also have an impact on  the BPS equation of this 
system and thus generalize the original Basu-Harvey equation \cite{Basu:2004ed}.

\item
Finally, it is a natural task to perform a similar analysis of superspace constraints
for the theories with less supersymmetry.
Of particular interest is the case ${\cal N}=6$, including the theories
of \cite{Aharony:2008ug,Hosomichi:2008jb}.
The relation to the harmonic superspace approach~\cite{Zupnik:2007he,Buchbinder:2008vi,Buchbinder:2009dc}
and the pure spinor formulations~\cite{Cederwall:2008xu} in this case remain to be investigated.
Also the question of a possible supersymmetry enhancement
from ${\cal N}=6$ to ${\cal N}=8$ may be addressed in this 
framework~\cite{Gustavsson:2009pm,Kwon:2009ar,Bandos:2009dt}.

\end{itemize}

We hope to come back to some of these issues in future work.
\bigskip

\noindent
{\bf Acknowledgements:}
This work is supported in part by the Agence Nationale de la Recherche (ANR).


\appendix

\section{A weaker constraint}\label{ap:WC}

In this appendix,
we complete the discussion of the constraint system (\ref{weakconstraint}),
i.e.\ of a vector multiplet 
with $W_{AB}$ considered as an independent field defined by (\ref{defW}). 
In this case, the constraint (\ref{defW}) can be understood as a partial flatness condition,
\begin{equation}
\label{flat}
\ca{F}_{\alpha A,\beta B}+ \ca{F}_{\alpha B,\beta A}=0\ ,
\end{equation}
and therefore admits an equivalent formulation as an linear auxiliary system,
\bea
\label{flat2}
\lambda^{\alpha\beta} \nabla_{A\beta} \,{\cal S}(\lambda) &=& 0\;,\qquad
\lambda^{\alpha\beta} \nabla_{\alpha\beta}\, {\cal S}(\lambda) ~=~ 0
\;,
\eea
with a light-like vector $\lambda^{\alpha\beta}\lambda_{\alpha\beta}=0$, such that integrability of (\ref{flat2}) implies 
(\ref{flat}). Light-like vectors in $\mathbb{R}^{1,2}$ are parametrized by $TS^1$, the Minkowski 
space version of the mini-twistor 
space \cite{Chiou:2005jn}, 
which suggest the existence of a corresponding twistor space formulation of this system. 

To keep the analysis of the multiplet structure transparent we analyze the system 
(\ref{defW}) for abelian vector superfields, for which the resulting equations simplify considerably.
The full non-abelian analysis does not add any conceptual challenges or modifications of the component
field content except for the fact that all fields are matrices of the non abelian Lie algebra.

The conditions due to the Bianchi identities (\ref{constraint160}),(\ref{FL}), (\ref{Dlambdarho}) and (\ref{eq:2.10}) are of the same form as in the non-abelian case, except that 
the covariant derivatives acting in the adjoint representation can be replaced by partial derivatives in the abelian case.  
The integrability conditions (\ref{icNA})  are  now
genuine nontrivial conditions on the superfields. In the abelian case, they simplify considerably to
\csep
\begin{eqnarray}
\label{icAB}
D_{\alpha A}\, \rho_{\beta BCD} &=&
     3i\del_{\alpha\beta} W_{[BC}\delta_{D]A}
         -\ft{3i}2\vep_{\alpha\beta}\delta_{A[B}V_{CD]}
           +iU_{\alpha\beta\, ABCD}\ ,\nonumber\\                          
  D_{\alpha A} V_{BC}  &=&  2\vep^{\beta\gamma} \del_{\alpha\beta}\left(
                              \delta_{A[B}\lambda_{C]\gamma}-\rho_{\gamma ABC}\right) \ ,\nonumber\\      
 D_{\alpha A} U_{\beta\gamma\, BCDE}&=&
                          8\delta^{A[B} \del_{\alpha(\beta}\rho_{\gamma)}^{CDE]}
                           -4\delta^{A[B} \del_{\beta\gamma}\rho_{\alpha}^{CDE]}
                             +\tau_{\alpha\beta\gamma\, ABCDE}\ .
\end{eqnarray}
Evaluating the anti-commutator (\ref{defW}) on the last equation of (\ref{icAB}) determines
the superderivative of the tensor $\tau_{\alpha\beta\gamma\, ABCDE}$ as
\bea
D_{\alpha A} \tau_{\beta_1\beta_2\beta_3\, B_1\cdots B_5} &=&
10i\delta^{A[B_1}\del_{\alpha(\beta_1}U^{B_2\cdots B_5]}_{\beta_2\beta_3)}
-5i\delta^{A[B_1}\del_{(\beta_1\beta_2}U^{B_2\cdots B_5]}_{\beta_3)\alpha}
\nonumber\\
&&{}
+iT_{\alpha\beta_1\beta_2\beta_3\, AB_1\cdots B_5}
\;,
\label{DT}
\eea
up to a tensor $T_{\alpha_1\cdots \alpha_4\, A_1 \cdots A_6}=
T_{(\alpha_1\cdots \alpha_4)\, [A_1 \cdots A_6]}$\,.
Iterating this procedure, we finally arrive at the (closed) system
\csep
\begin{eqnarray}
\label{DTSS}
D_{\alpha A} T_{\beta_1\cdots \beta_4\, B_1\cdots B_6} &=&
12\,\delta^{A[B_1}\del_{\alpha(\beta_1}T^{B_2\cdots B_6]}_{\beta_2\beta_3\beta_4)}
-6\,\delta^{A[B_1}\del_{(\beta_1\beta_2}T^{B_2\cdots B_6]}_{\beta_3\beta_4)\alpha}
\nonumber\\
&&{}
+\sigma_{\alpha\beta_1\cdots \beta_4\, AB_1\cdots B_6}
\;,
\nonumber\\[2ex]
D_{\alpha A} \sigma_{\beta_1\cdots \beta_5\, B_1\cdots B_7} &=&
14i\delta^{A[B_1}\del_{\alpha(\beta_1}T^{B_2\cdots B_7]}_{\beta_2\cdots \beta_5)}
-7i\delta^{A[B_1}\del_{(\beta_1\beta_2}T^{B_2\cdots B_7]}_{\beta_3\beta_4\beta_5)\alpha}
\nonumber\\
&&{}
+iS_{\alpha\beta_1\cdots \beta_5\,AB_1\cdots B_7}
\;,
\nonumber\\[2ex]
D_{\alpha A} S_{\beta_1\cdots \beta_6\, B_1\cdots B_8} &=&
16\,\delta^{A[B_1}\del_{\alpha(\beta_1}\sigma^{B_2\cdots B_8]}_{\beta_2\cdots\beta_6)}
-8\,\delta^{A[B_1}\del_{(\beta_1\beta_2}\sigma^{B_2\cdots B_8]}_{\beta_3\cdots \beta_6)\alpha}
\;.
\end{eqnarray}
with additional tensors $\sigma$ and $S$, which are completely symmetric (antisymmetric) in their
$SO(2,1)$ ($SO(8)$) indices. 
Evaluating the anti-commutator (\ref{defW}) on the first equation of (\ref{icAB}) 
leads to two consistency equations for the tensor  and $U_{ABCD\,\alpha\beta}$ and the fourth (abelian) Bianchi identity:
\bea
\del^{\alpha\beta}\ca{F}_{\alpha\beta} &=& 0\;,
\qquad
\del^{\alpha\beta}U_{\alpha\beta\,ABCD} ~=~ 0
\;,
\label{DDX}
\eea
Similarly, consistency of (\ref{DT}), (\ref{DTSS}) requires the first order equations
\bea
\del^{\alpha\beta}\tau_{\alpha\beta\gamma\,A_1\cdots A_5} ~=~ 0
\;,
\qquad
\del^{\alpha\beta}T_{\alpha\beta\gamma_1\gamma_2\,A_1\cdots A_6\,} ~=~ 0
\;,
\label{DDT}
\eea
and analogous equations for $\sigma$ and $S$,  showing that in the abelian case these tensors are conserved higher spin currents.
In the non-abelian case, a crucial modification takes place. First, partial derivatives are replaced by covariant derivatives 
and second,  the r.h.s.\ of the equations (\ref{DDX}), (\ref{DDT}) (except for the Bianchi identity) receive non-vanishing contributions from commutators of the non-abelian fields.

\subsubsection*{Superfield expansion, multiplet structure}

The obtained closed system 
of superderivatives of superfields (\ref{defW}), (\ref{constraint160}), (\ref{FL}), 
(\ref{Dlambdarho}), (\ref{eq:2.10}) and  (\ref{icAB}), (\ref{DT}), (\ref{DTSS}) allows to 
define a closed recursive
system to systematically obtain the expansion in terms of component fields. Contracting all
these equations  with $\theta^{\alpha A}$ gives
\csep
\begin{eqnarray}
\label{A7}
(1+{\cal R})\, \ca{A}_{\alpha A} &=& 2i\theta^{\beta A}\ca{A}_{\alpha\beta}
    +2i\vep_{\alpha\beta}\theta^{\beta B} W_{AB} \;,\nonumber\\
{\cal R}\, \ca{A}_{\alpha\beta} &=& \theta^{\gamma A} \vep_{\gamma(\alpha} \lambda_{\beta)A}
\;,\nonumber\\
{\cal R}\, W_{AB} &=& \theta^{\delta D}(\delta_{D[A}\lambda_{B]\delta} + \rho_{\delta DAB})
\;,\nonumber\\
\ca{R}\, \lambda_{\alpha A} &=&
i\theta^{\delta D} (\delta_{DA}\ca{F}_{\delta\alpha}+ 2\  \del_{\delta\alpha} W_{DA}
 +\vep_{\delta\alpha}V_{DA})\;, \nonumber\\
&\cdots&\nonumber\\
{\cal R}\,  S_{\alpha_1\cdots \alpha_6\,A_1\cdots A_8} &=&
16\,\theta^{\beta[A_1}\del_{\beta(\alpha_1}\sigma^{A_2\cdots A_8]}_{\alpha_2\cdots\alpha_6)}
-8\,\theta^{\beta[A_1}\del_{(\alpha_1\alpha_2}\sigma^{A_2\cdots A_8]}_{\alpha_3\cdots \alpha_6)\beta}
\;,
\end{eqnarray}
generalizing (\ref{eq:1.29}).
This shows that the superfield $\ca{A}_{\alpha A}$ is entirely determined in terms of the 
lowest components of all the superfields involved
\begin{eqnarray}
\label{AaA}
\ca{A}_{\beta B}&=&i (\theta^{\alpha}_{B}\ A_{\alpha\beta} + \theta^{\alpha A}\vep_{\alpha\beta}\ov{W}_{AB})
            \nonumber\\ &&\ \  +\fr{2i}{3}\, \theta^{\alpha A}\theta^{\gamma C}
    (\delta_{AB}\ \vep_{\gamma(\alpha}\ov{\lambda}_{\beta)C}
      +\vep_{\alpha\beta}\delta_{C[A}\ov{\lambda}_{B]\gamma}
                    +\vep_{\alpha\beta}\ \ov{\rho}_{\gamma CAB})+\ldots\ .
\end{eqnarray}

The only equations that these fields must obey are the first order constraint equations
(\ref{DDX}), (\ref{DDT}), etc.
The superfield expansion of $\ca{A}_{A\alpha}$ is summarized in table~\ref{tab:Aa},
where the negative multiplicities refer to the first order constraint equations.
The resulting multiplet is thus neither on-shell (as there are genuine field equations for its components) nor 
entirely off-shell (due to the presence of
the constraint equations).
Counting the field content of table~\ref{tab:Aa} reveals $257$ bosonic $+$ $256$ fermionic degrees of freedom
with the extra bosonic singlet corresponding to the gauge freedom of the
vector field $A_{\alpha\beta}$.
Interestingly, the same multiplet has
appeared in~\cite{Gervais:1999vj} in the context of reducing
the superspace constraints of ten-dimensional Yang-Mills theories
down to seven dimensions.

\begin{TABLE}[bt]
{\begin{tabular}{c|c|c}
$\theta^N$ & Field & Representation under $SO(2,1)\times SO(8)$\\
\hline
$0$ & --- & ---\\
$1$ & $A_{\alpha\beta}+ \ov{W}_{AB}$ & 
${\bf (3,1)}+{\bf (1,28)}$
\\
$2$ & $ \ov{\lambda}_{\alpha A}+ \ov{\rho}_{\alpha\,ABC}$ & 
           ${\bf (2,8_s)}+{\bf (2,56_s)}$\\
$3$ & $
\ov{V}_{AB}+ \ov{U}_{\alpha\beta\,ABCD}$ & ${\bf (1,28)}+{\bf (3-1,35_v+35_c)}$\\
$4$ & $ \ov{\tau}_{\alpha\beta\gamma\,ABCDE}$ & ${\bf (4-2,56_s)}$ \\
$5$ & $ \ov{T}_{\alpha_1\dots\alpha_4\,A_1\dots A_6}$ & ${\bf (5-3,28)}$ \\
$6$ & $ \ov{\sigma}_{\alpha_1\dots\alpha_5\,A_1\dots A_7}$ & ${\bf (6-4,8_s)}$ \\
$7$ & $ \ov{S}_{\alpha_1\dots\alpha_6\,A_1\dots A_8}$ & ${\bf (7-5,1)}$ \\
$8$ & --- & ---\\
\end{tabular}
\caption{
Superfield expansion of the vector field $\ca{A}_{\alpha A}$ induced by the 
weaker constraint (\ref{weakconstraint}). The negative multiplicities of 
representations w.r.t. $SO(2,1)$ correspond to the first order constraint equations which these fields satisfy.}
\label{tab:Aa}}
\end{TABLE}

The relation between $F_{\alpha\beta}$ and $A_{\alpha\beta}$ may give an
idea how to resolve the constrained fields in terms of genuine off-shell fields.
E.g.\ in the abelian theory, the 70 conserved currents $U_{ABCD\,\alpha\beta}$ can be written
in the form
\bea
U^{ABCD}{}_{\alpha\beta} &=& 
\vep^{\gamma\delta} \partial_{\gamma(\alpha} B^{ABCD}{}_{\beta)\delta}
\;,
\eea
as the field strengths of 70 off-shell and unconstrained vector fields $B^{ABCD}{}_{\alpha\beta}$\,.
For the higher spin fields in contrast, this is less clear. In particular, 
the non-abelian generalization
upon which the components $U^{ABCD}{}_{\alpha\beta}$, 
$\tau_{\alpha\beta\gamma\,A_1\cdots A_5}$, etc., are no longer 
covariantly conserved currents,  makes it even harder to see 
if there exists an formulation in terms of genuine off-shell fields. 

In the non-abelian case the superfield expansion of $\ca{A}_{\alpha A}$ to second order in $\theta$  is formally
the same as in (\ref{AaA}). For the basic matter superfields and the bosonic gauge superfield one finds to second 
order in $\theta$:
\begin{eqnarray}
\label{eq:psiexp}
\Phi^I&=&\phi^I+i \theta^{\alpha A}\Gamma^I_{A\D{A}} \psi_{\alpha \D{A}}
+\fr{i}{2}\, \theta^{\alpha A}\theta^{\beta B}\ \Gamma^{IJ}_{AB}\  \ov{\nabla}_{\alpha\beta}\phi^J
\nonumber\\ &&\ \  -\fr{i}{4}\, \theta^{\alpha A}\theta^{\beta B}\vep_{\alpha\beta}
     ( \fr{1}{7}\, \delta_{AB}\  \ov{W}_{IJ}\cdot\phi^J-\fr{1}{6}\, \Gamma^{ILMN}_{AB}\ \ov{W}_{LM}\cdot\phi_{N})
+\ldots \ ,\nonumber\\[.5ex]
\Psi_{\beta\D{A}}&=&\psi_{\beta\D{A}}+\theta^{\alpha A}(\Gamma^I_{A\D{A}}\ov{\nabla}_{\alpha\beta}\phi^I+
\hal\ \vep_{\alpha\beta}P^{IJK}_{A\D{A}}\ \ov{W}_{IJ}\cdot\phi_K)\nonumber\\
&& +\fr{i}{2}\ \theta^{\alpha A}\theta^{\gamma C}
 \Big(\Gamma^I_{A\D{A}}\Gamma^I_{C\D{C}}\ov{\nabla}_{\alpha\beta}\  \psi_{\gamma \D{C}} 
 +P^{IJK}_{A\D{A}}\Gamma^K_{C\D{C}}\ \ov{W}_{IJ}\cdot\psi_{\gamma\D{C}}\Big)\nonumber\\
 && +\fr{1}{2}\, \theta^{\alpha A}\theta^{\gamma C}
 \Big(\Gamma^K_{A\D{A}}\ \vep_{\gamma(\alpha}\ov{\lambda}_{\beta)C}\cdot\phi^K
     +\fr{1}{4}\, \Gamma^{IJ}_{BD}P^{IJK}_{A\D{A}}(\delta_{C[B}\ov{\lambda}_{D]\gamma}
   +\ov{\rho}_{\gamma CBD})\cdot\phi_K\Big)+\ldots\ , \nonumber\\[.5ex]
\ca{A}_{\alpha\beta}&=& A_{\alpha\beta}
  +\theta^{\gamma C}\vep_{\gamma(\alpha}\ov{\lambda}_{\beta)C} \nonumber\\
&& +\fr{i}{2}\, \theta^{\gamma C}\theta^{\delta D}[\hal \vep_{\gamma\delta}\ \delta_{CD} F_{\alpha\beta}
-2\vep_{\gamma(\alpha}\ov{\nabla}_{\beta)\delta}\ov{W}_{CD}
    +\vep_{\gamma(\alpha}\vep_{\beta)\delta}\ov{V}_{CD}]+\ldots\ ,
\end{eqnarray}
where we have introduced the abbreviation 
$P^{IJK}_{A\D{A}}= \fr{1}{7}\, \Gamma^I_{A\D{A}}\delta^{JK}+\fr{1}{6}\, \Gamma^{IJK}_{A\D{A}}$.

\section{$SO(8)$ relations}\label{ap:gam}

The group $SO(8)$  (we mainly consider the associated Lie-algebra $so(8)$ and we 
are somewhat cavalier regarding the difference)
has rather special properties. It admits a Majorana-Weyl 
representation in terms of real eight-component Spinors and the chirally conjugated ones, and consequently there are 
three inequivalent (real) eight-dimensional irreducible 
representations ${\bf{8_s}}$,  ${\bf{8_c}}$ and  ${\bf{8_v}}$,
where ${\bf{8_v}}$ is the vector representation of $SO(8)$. The source of this 
``accidental'' coincidence in the dimensionality
is the underlying triality symmetry which can be seen from the associated Dynkin diagram.

A commonly chosen Majorana-Weyl representation of the $SO(8)$ Gamma matrices $\tilde{\Gamma}^I$ is given in 
terms of real $8\times 8$ blocks:
\begin{equation}
\label{eq:a2.1}
\tilde{\Gamma}^I=\left [\begin{array}{cc}0&\Gamma^I\\\bar{\Gamma}^I&0\end{array}\right]\ ,
\end{equation}
where $\bar{\Gamma}^I=(\Gamma^I)^T$. We denote the components of the matrices $\Gamma^I$ by
\begin{equation}
\label{gam}
\Gamma^I_{A\D{B}}\quad\quad\mathrm{with}\quad\quad I, A, \D{B}=1,\ldots, 8\ \ ,
\end{equation}
and we do not introduce a separate symbol for the transposed matrices $\bar{\Gamma}^I$, which in fact 
occur only in this appendix to keep the notation more compact. The basic algebraic relations for these 
matrices are\footnote{We denote symmetrization/antisymmetrization in indices by $()$ and $[]$, respectively, and 
(anti)-symmetrizations are always defined with weight one.}
\begin{equation}
\label{eq:14}
\Gamma^{(I}\bar{\Gamma}^{J)}=\bar{\Gamma}^{(I}\Gamma^{J)}=\delta^{IJ}\  \unit_{8} \ \ ,
\end{equation}
and an explicit representation of these matrices can be found for example in \cite{Green:1987sp}. Further 
we introduce the totally antisymmetrized products
\begin{eqnarray}
\label{eq:15}
\Gamma^{I_1I_2\ldots I_n}_{AB}:=(\Gamma^{[I_1}\bar{\Gamma}^{I_2}\ldots\bar{\Gamma}^{I_n]})_{AB}\ \ 
\ldots&& n \ \mathrm{even} \ , 
\nonumber\\ 
\Gamma^{I_1I_2\ldots I_n}_{A\D{A}}:=(\Gamma^{[I_1}\bar{\Gamma}^{I_2}\ldots\Gamma^{I_n]})_{A\D{A}}
   \ \ \ldots&& n\  \mathrm{odd} \ ,
\end{eqnarray}
and analogously one can define matrices $\bar{\Gamma}^{IJK\ldots}$ where the alternating sequence of matrix products
starts with a transposed matrix $\bar{\Gamma}^I$, replacing dotted and undotted indices in (\ref{eq:15}). 
These matrices have the following symmetry properties under transposition:
\begin{eqnarray}
\label{eq:16}
\Gamma^{I_1I_2\ldots I_n}_{AB}=(-)^{n(n-1)/2}\  \Gamma^{I_1I_2\ldots I_n}_{BA}\ \ 
      \ldots&& n \ \mathrm{even}\ ,\nonumber\\
( \Gamma^{I_1I_2\ldots I_n})^T=(-)^{n(n-1)/2}\  \bar{\Gamma}^{I_1I_2\ldots I_n}
 \ \ \ldots&& n\  \mathrm{odd} \ .
\end{eqnarray}

{\bf{Identities.}} We give here a number of useful  $\Gamma$-matrix identities 
which where used in the calculations 
of the main text. We first give a basic identity, which is also the origin of the triality relations that
we used in this work (see below): 
\begin{equation}
\label{eq:17}
\Gamma^I_{A\D{A}}\Gamma^I_{B\D{B}}+\Gamma^I_{A\D{B}}\Gamma^I_{B\D{A}}=2\ \delta_{AB}\delta_{\D{A}\D{B}}\ \ .
\end{equation}
Defining $\delta^{I_1\ldots I_n}_{J_1\ldots J_n}:=\delta^{I_1}_{[J_1}\ldots \delta^{I_n}_{J_n]}$  we have the following identities:
\begin{itemize}
\item Traces
\begin{eqnarray}
\label{eq:19}
{\rm Tr}[\Gamma^{I_1\ldots I_n}]&=&0\ \ \mathrm{for}\ \ n>1\;,\nonumber\\
{\rm Tr}[\Gamma^{IJ}\Gamma^{KL}]&=&-16 \, \delta^{IJ}_{KL} \;,\nonumber\\
{\rm Tr}[\Gamma^{IJK}\Gamma^{LMN}]&=&48 \, \delta^{IJK}_{LMN} \;,\nonumber\\
{\rm Tr}[\Gamma^{IJKL}\Gamma^{MNOP}]&=&8 \,(24 \, \delta^{IJKL}_{MNOP}+\vep^{IJKLMNOP}) \;,\nonumber\\
{\rm Tr}[\Gamma^{IL}\Gamma^{JM}\Gamma^{KN}]&=&32 \, 
(\delta_{L}^{[J}\delta_{KN}^{M]I}-\delta_I^{[J}\delta_{KN}^{M]L})
\;.
\end{eqnarray}
\item Products
\csep
\begin{eqnarray}
\label{eq:20}
(\Gamma^{IJ}\Gamma^{KL})_{AB}&=&\Gamma^{IJKL}
      -2\, (\delta^{K[I}\Gamma^{J]L}_{AB}-\delta^{L[I}\Gamma^{J]K}_{AB})
      -2\, \delta_{AB}\ \delta^{IJ}_{KL}\;,\nonumber\\
(\Gamma^{LN}\Gamma^{IJKN})_{AB}&=&4\ \Gamma^{IJKL}_{AB}-30\, \Gamma^{[IJ}_{AB}\delta^{K]L}
\;,\nonumber\\
(\bar{\Gamma}^{ILN}\Gamma^{IJK})_{\D{A}\D{B}}&=& 
4\, \bar{\Gamma}^{LNJK}_{\D{A}\D{B}}+10\, (\bar{\Gamma}^{L[K}_{\D{A}\D{B}}\delta^{J]N}
           -\bar{\Gamma}^{N[K}_{\D{A}\D{B}}\delta^{J]L})+12\, \delta^{LN}_{KJ}\delta_{\D{A}\D{B}}
           \;.
\end{eqnarray}
\item Tensor products
\csep
\begin{eqnarray}
\label{eq:21}
\Gamma^{IJ}_{AB} \Gamma^{IJ}_{CD}&=&16 \, \delta^{AB}_{CD} \;,\nonumber\\
\Gamma^{IJ}_{AB} \bar{\Gamma}^{IJ}_{\D{C}\D{D}}&=&
2 \, \Gamma^I_{A\D{C}}\Gamma^I_{B\D{D}}
-2 \, \Gamma^I_{B\D{C}}\Gamma^I_{A\D{D}}
\;, \nonumber\\
\Gamma^{IJKL}_{AB}\Gamma^L_{C\D{C}}&=&-\delta_{AB}\Gamma^{IJK}_{C\D{C}}
      +2  \,\delta_{C(A}\Gamma^{IJK}_{B)\D{C}} + 6 \,\Gamma^{[IJ}_{C(A}\Gamma^{K]}_{B)\D{C}}\;,\nonumber\\
\Gamma^{IJ}_{AB}\Gamma^{IJK}_{C\D{C}}&=&-2 \,\Gamma^I_{C\D{C}}\Gamma^{IK}_{AB}
             +16 \,\delta_{C[A}\Gamma^K_{B]\D{C}}\;,\nonumber\\
(\bar{\Gamma}^{J}\Gamma^{I}\bar{\Gamma}^K)_{\D{A}A}\Gamma^{JK}_{BC}&=&
        16 \,\delta_{A[B}\Gamma^I_{C]\D{A}}
              -2 \,\Gamma^J_{A\D{A}}\Gamma^{JI}_{BC}\;,\nonumber\\
\Gamma^{IJ}_{AB}\Gamma^{IJKL}_{CD}&=&2 \,\delta_{CD}\Gamma^{KL}_{AB} 
             - 8 \,(\delta_{A(C}\Gamma^{KL}_{D)B}-\delta_{B(C}\Gamma^{KL}_{D)A})\;,\nonumber\\
\Gamma^{IJK}_{A\D{A}}\Gamma^{IJK}_{B\D{B}}&=&48 \,\delta_{AB}\delta_{\D{A}\D{B}}
         -6 \,\Gamma^I_{A\D{A}}\Gamma^I_{B\D{B}}\;,\nonumber\\
\Gamma^{IJK}_{A\D{A}}\Gamma^{IJKL}_{BC}&=&48 \,\delta_{A(B}\Gamma^L_{C)\D{A}}
                                   -6 \,\delta_{BC}\Gamma^L_{A\D{A}}
                                   \;.
\end{eqnarray}
\end{itemize}

{\bf{Triality.}}  Here we explain some triality relations which were
used in the main text. The  basic identity for these considerations is equation (\ref{eq:17}) 
which is exactly the same relation as (\ref{eq:14}) if we consider ``new'' matrices\footnote{We do not introduce 
a new symbol for these matrices but take the index name from the range $A,B,C..$ as opposed 
to $I,J,K...$ as part of the defining symbol, in particular this means for example $\Gamma^{A=1}\neq\Gamma^{I=1}$.}
$\Gamma^{A}$ with matrix components $\Gamma^A_{I\D{B}}:=\Gamma^I_{A\D{B}}$  (the same is true for matrices 
$\Gamma^{\D{B}}$ with matrix components $\Gamma^{\D{B}}_{AI}:=\Gamma^I_{A\D{B}}$). Thus the 
matrices $\Gamma^A$ provide the same algebraic structure as the matrices 
$\Gamma^I$ and we can define the analogous antisymmetrized products $\Gamma^{ABCD\ldots}$ 
as in (\ref{eq:15}) with the same properties and analogous formulas as in (\ref{eq:19}), (\ref{eq:20}) and (\ref{eq:21}) 
will hold for them. 
In addition we can reinterpret different expressions in the tensor products (\ref{eq:21}). A particular example that was used 
in the main text is:
\begin{equation}
\label{tri}
\Gamma^{I}_{C\D{C}}\Gamma^{IJ}_{AB}=\Gamma^{C}_{I\D{C}}\Gamma^{AB}_{IJ}
 =-(\Gamma^{AB}\bar{\Gamma}^C)_{J\D{C}}
=-(\Gamma^{ABC}_{J\D{C}}+2\ \Gamma^{[A}_{J\D{C}}\delta^{B]C})\ ,
\end{equation}
with $\Gamma^{ABC}_{I\D{C}}\equiv  \Gamma^{IJ}_{[AB}\Gamma^{J}_{C]}{}_{\D{C}}$.
In the main text we also use the fact that the adjoint
representation of $so(8)$ can be written as
\bea
{\bf 28} = ({\bf 8_v} \otimes {\bf 8_v})_{\rm alt} = 
({\bf 8_s} \otimes {\bf 8_s})_{\rm alt} = 
({\bf 8_c} \otimes {\bf 8_c})_{\rm alt} \;,
\eea
which allows to label tensors in this representation by different antisymmetric index pairs, e.g.
\begin{equation}
\label{tri1}
W_{IJ}\equiv\fr{1}{4} \Gamma^{IJ}_{AB}W_{AB}\quad  , \quad
W_{AB}\equiv\fr{1}{4} \Gamma^{IJ}_{AB}W_{IJ}\quad , \quad
W_{\D{A}\D{B}}\equiv\fr{1}{4}\bar{\Gamma}^{IJ}_{\D{A}\D{B}}W_{IJ}
  \ ,\quad {\rm etc.}\ .
\end{equation}

\section{$SO(2,1)$ spinor conventions}

All spinors appearing in the main text, superspace coordinates or fields,  are Majorana spinors in $2+1$-dimensional 
space-time. Our  metric convention is
$\eta_{\mu\nu}=(-,+,+)$ and we choose a Majorana representation for the 
gamma-matrices\footnote{In terms of the Pauli matrices $\sigma^i$ for example 
$\gamma^0=-i\sigma^2\ ,\gamma^1=\sigma^1 , \gamma^2= \sigma^3$, see e.g.~\cite{RuizRuiz:1996mm} 
for more details.}
\begin{equation}
\label{eq:a1}
\{\gamma^\mu,\gamma^\nu\}^\alpha{}_{\beta}=2\eta^{\mu\nu}\delta^\alpha{}_{\beta}\ \ .
\end{equation}
Thus the matrices $\gamma^{\mu\ \alpha}{}_{\beta}$ are real and the Majorana condition on spinors imply 
that they are real two component spinors. Spinor indices are raised/lowered by the epsilon symbols with 
$\vep^{12}=\vep_{12}=1$ and choosing NW-SE conventions
\begin{equation}
\label{eq:a2}
\vep^{\alpha\gamma}\vep_{\beta\gamma}=\delta^\alpha_{\ \beta}\  ,\quad 
\lambda^\alpha:=\vep^{\alpha\beta}\lambda_\beta\Leftrightarrow\lambda_\beta 
    = \lambda^\alpha\vep_{\alpha\beta}\ .
\end{equation}
Introducing the real symmetric matrices 
$\sigma^\mu_{\alpha\beta}:=\gamma^{\mu\ \rho}{}_{\beta}\  \vep_{\rho\alpha}$ and 
$\bar{\sigma}^{\mu\ \alpha\beta}:= (\vep \cdot \sigma^\mu \cdot \vep)^{\alpha\beta}
=-\vep^{\beta\rho}\  \gamma^{\mu\ \alpha}{}_{\rho} $ a three vector in spinor notation writes as 
a symmetric real matrix as 
\begin{equation}
\label{eq:a3}
v_{\alpha\beta}:=\sigma^\mu_{\alpha\beta}\ v_\mu\ \ \Rightarrow \ \ 
 v^\mu=\fr{1}{2}\bar{\sigma}^{\mu\ \alpha\beta}\ v_{\alpha\beta}\ .
\end{equation}

\bibliographystyle{JHEP}

\providecommand{\href}[2]{#2}\begingroup\raggedright\endgroup

\end{document}